\documentclass[aip,cha,reprint]{revtex4-1}

\usepackage{epsfig}
\usepackage{float}
\usepackage{verbatim}
\usepackage{amssymb}
\usepackage{bbold}
\usepackage{amsmath}
\usepackage{psfrag}
\usepackage{color}
\usepackage{ulem}

\usepackage[latin1]{inputenc}
\usepackage[english]{babel}
\usepackage[colorlinks, citecolor=blue, linkcolor=red, urlcolor=blue]{hyperref}
\usepackage{graphicx}
\usepackage{amscd}
\usepackage{eucal}
\usepackage{bm}
\usepackage{lipsum}
\usepackage{url}

\input {epsf}
\begin{document}

\title{Route to Chaos in Multi-Species Ecosystems}

\author{R. Delabays}
\email{robin.delabays@hevs.ch}
\affiliation{School of Engineering, University of Applied Sciences of Western Switzerland HES-SO, CH-1950 Sion, Switzerland}
\author{Ph. Jacquod}
\email{philippe.jacquod@unige.ch}
\affiliation{School of Engineering, University of Applied Sciences of Western Switzerland HES-SO, CH-1950 Sion, Switzerland}
\affiliation{Andlinger Center for Energy and the Environment, Princeton University, Princeton, NJ 08544 USA}
\affiliation{Department of Quantum Matter Physics, University of Geneva, CH-1211 Geneva, Switzerland}

\date{\today}

\begin{abstract}
We investigate species-rich mathematical models of ecosystems. 
While much of the existing literature focuses on the properties of equilibrium fixed points, persistent dynamics (e.g., limit cycles or chaos) have also been observed, both in natural or lab-controlled ecosystems
and in mathematical models. 
Here we emphasize the emergence of limit cycles following Hopf bifurcations tuned by the variability of interspecies interaction. 
As this variability increases, and owing to the large dimensionality of the system, limit cycles typically acquire a growing spectrum of frequencies. 
This often leads to the appearance of strange attractors, with a chaotic dynamics of species abundances characterized by a positive Lyapunov exponent. 
We observe that limit cycles and strange attractors preserve biodiversity to some extent, as they maintain dynamical stability without species extinction.
We give numerical evidences that this route to chaos dominates in ecosystems with strong enough interactions and where predator-prey behavior dominates over competition and mutualism. 
Based on arguments from random matrix theory, we further conjecture that this scenario is generic in ecosystems with large number of species, and identify the key parameters driving it. 
Overall, we show that the model we consider provides a unifying framework, where a wide range of population dynamics emerge from a simple few-parameter model. 
\end{abstract}

\maketitle

\begin{quotation}
 The time evolution of a species population in a given ecosystem can follow very diverse paths. 
 Of course, a population can remain relatively stable over time, but it can also oscillate between highs and lows, from one year to another. 
 More surprisingly, recent studies have shown that this evolution can even be chaotic -- meaning that no regular patterns emerge, and the population at any given time is highly sensitive to its initial state.
 We show that the three behaviors aforementioned (steady state, cycle, and chaos) can be modeled in a parcimonious version of the generalized Lotka-Volterra model. 
 Furthermore, we identify the two model parameters that govern the transition between these three states. 
 Namely, we observe that, as the interaction strengths between species increases, ecosystems tend to transition from a steady state, to a cyclic trajectory, and further to a chaotic behavior.
\end{quotation}

\section{Introduction}
One of the main challenges in theoretical ecology is to connect predictions from mathematical models of population dynamics to 
empirical observations of species coexistence in natural or laboratory-controlled ecosystems.~\cite{May07} 
In the latter it is established that individual populations fluctuate in time, often with large qualitative and quantitative differences between species.~\cite{Ran98,Lun00,Inc03,Van06} 
In some instances, population abundances exhibit synchronized, periodic oscillations, while in others population dynamics appears chaotic.~\cite{Has93,Ell95,Hun98,Hui99,Ben08,Rog22,Pre10,Pea20}
It is highly desirable to find out whether the wide range of observed dynamics can be captured in a single mathematical model, by varying few parameters, and if yes, what are the key characteristics this model should retain. 
Investigating mathematical models can furthermore shed light on fundamental qualitative questions such as whether populations fluctuations are endogenous or exogenous, i.e., if they are generated by intrinsic interactions or by external sources.~\cite{Ell95,Sch03}

In this manuscript, we numerically investigate large Lotka-Volterra models with random interactions, which are standard multi-species models of population dynamics. 
We emphasize the richness of their dynamics as a function of two key parameters which are  (i) the variability $\sigma$ of interspecies interactions, and (ii) the cross-diagonal covariance parameter $\gamma$ of the interaction matrix. 
Our main finding is that, for a predominance of predator-prey pairs of species, the stable fixed-points prevailing at weak interaction variability generically lose their stability through Hopf bifurcations. 
Limit cycles emerge, where surviving species have periodically oscillating abundances. 
At still stronger interactions, strange attractors appear, possibly from cascades of bifurcations, which lead to a chaotic dynamics of population abundances characterized by a positive largest Lyapunov exponent. 
This route to chaos illustrates how stationarity, oscillating periodicity and chaos in the  dynamics of species abundances exist in rather general models of population dynamics, depending on $\sigma$ and $\gamma$.
One important result is that all observed population dynamics in multi-species ecosystems can be reproduced by a unified and parcimonious mathematical model.

Much of the existing literature on theoretical ecology is based on the surmise that the observed states and dynamics of ecosystems can be described by the time-asymptotic behavior of mathematical models. 
The latter is characterized by either a time-independent, stationary distribution of population abundances, or by persistently varying abundances, oscillating periodically or aperiodically. 
Accordingly, there has been a significant focus on the equilibrium fixed points of large systems of competing species, their stability and feasibility.~\cite{May72,Rob74,Rie89,All12,Tan14,Fyo16,Cle23,Ros23}
One key parameter is the variability $\sigma$ of the interspecies interaction. 
It has been found that a unique, asymptotically stable fixed point exists at weak enough $\sigma<\sigma_c$,~\cite{May72,Rie89,Cle23,Ros23} and that $\sigma_c$ increases in ecosystems with dominating predator-prey interactions.~\cite{All12,Tan14}
For $\sigma > \sigma_c$, one enters a phase with multiple unstable equilibria.~\cite{Ros23} 
Stabilization may still occur via species extinction, which drives the ecosystem to a novel, stable fixed point with reduced biodiversity. 
Beyond fixed points, the emergence of limit cycles through a Hopf bifurcation has been emphasized, albeit for small systems with only few species.~\cite{Gil75,Cos79,Gar89,Fus00,Mcg08}
Further regimes with aperiodic persistent dynamics have recently been highlighted at stronger interactions and/or larger number of species,~\cite{Hu22} in ecosystems with sparse interactions,~\cite{Roy20,Mar24} or communities with migrations.~\cite{Dal20} 

In contrast to this mainstream philosophy focusing on the time-asymptotic behavior of mathematical models, several works have advocated that sufficiently large ecosystems have long relaxation times so that, by nature, they are always observed in a transient state.~\cite{Has94,Schr03,Mor16,Has18}
Such large relaxation times may follow for instance from the slowing down of the dynamics in ecosystems close to criticality.~\cite{Sol02,Mor11} 
Below we show that all these different behaviors naturally emerge in different regimes of a single mathematical model, as a function of only two parameters. 

\section{Model and method} 
Population dynamics in multi-species ecosystems is commonly studied in the framework of the generalized Lotka-Volterra model.~\cite{Goe71} 
Here we consider the variant investigated in Ref.~\onlinecite{Ros23}, which reads
\begin{equation}\label{lv}
 \dot N_i = N_i \left[ \kappa_i- N_i -\frac{\mu}{S} \sum_{j=1}^S N_j - \frac{\sigma}{\sqrt{S}} \sum_{j=1}^S {\mathbb A}_{ij} N_j \right] \, .
\end{equation}
Equation~\eqref{lv} determines the time-evolution of the normalized abundance $N_i(t) \ge 0$ of species $i=1,2, \ldots S$, as a function of its intrinsic growth rate $\kappa_i$ and its interaction with other species. 
Interspecies interactions have a finite average $\mu/S$ and a variability $\sigma/\sqrt{S}$. 
The chosen scaling of interaction strengths with $S$ guarantees that the spectral support of the stability matrix [see Eq.~\eqref{matrel} below] does not change with $S$.
Formulating the system as in Eq.~\eqref{lv} emphasizes the distribution underlying the random inter-species interactions. 
It allows us to pinpoint the relevant parameters in the route to chaos.

Fluctuations in interaction strengths between pairs of species are encoded in the components ${\mathbb A}_{ij}$ of the interaction matrix $\mathbb{A}$. 
Being interested in large, heterogeneous ecosystems with no particular structure, we follow a random matrix theory (RMT) approach,~\cite{May72,All15,Akj24} where the matrix $\mathbb A$ belongs to an ensemble of matrices with elements ${\mathbb A}_{ij}$ that are normally distributed with vanishing ensemble average,  $\langle {\mathbb A}_{ij}\rangle =0$, and covariances  given by 
\begin{equation} 
\label{randommatr}
\langle {\mathbb A}_{ij} {\mathbb A}_{kl} \rangle = \delta_{ik} \delta_{jl}+\gamma \delta_{il} \delta_{jk} \, .
\end{equation}
The cross-diagonal covariance parameter $\gamma \in [-1,1]$ allows to tune the types of two-species interactions that are likely to be encountered in the model. 
For $\gamma=1$, the interaction matrix ${\mathbb A}$ is totally symmetric and therefore interactions are either mutualistic (${\mathbb A}_{ij} \le 0$ and ${\mathbb A}_{ji} \le 0$), or competitive (${\mathbb A}_{ij} \ge 0$ and ${\mathbb A}_{ji} \ge 0$). 
For $\gamma=-1$, on the other hand, ${\mathbb A}$ is totally antisymmetric and all interactions are predator-prey (with ${\mathbb A}_{ij} \le 0$ and ${\mathbb A}_{ji} \ge 0$ of vice-versa).
Tuning $\gamma$ between $1$ and $-1$, one goes from a system dominated by mutualistic or competitive interaction ($\gamma$ close to $1$) to a system dominated by predator-prey interactions ($\gamma$ close to $-1$).~\cite{gamma} 
Even though the mean $\mu$ introduces a bias towards mutualistic/competitive interactions, its contribution becomes negligible with respect to the contribution of the variability $\sigma$ for large $S$, due to the respective scalings in $S$ of the corresponding terms in Eq.~\eqref{lv}. 
For $\gamma=0$, $\mathbb{A}$ belongs to the Ginibre ensemble or random matrices.~\cite{Gin65}

The generalized Lotka-Volterra model of Eq.~\eqref{lv} assumes that $N_i(t)$ is real and varies continuously. 
This is a legitimate assumption only as long as $N_i(t)$ is sufficiently large. 
Large fluctuations have been observed in numerical simulations of Eq.~\eqref{lv}, where some species resurrect from minuscule abundances, effectively corresponding to extinction. 
As a matter of fact, the dynamics of Eq.~\eqref{lv}  leads to species extinction only for asymptotically long times, and the standard procedure to solve this {\it atto-fox} problem~\cite{Mol91} is to introduce a small, but finite extinction threshold $N_c$. 
When $N_i(t_c) < N_c$, extinction occurs and $N_i(t>t_c)\equiv 0$. 
We will set $N_c=10^{-20}$ but have checked that other choices lead to the same conclusions as presented below. 

The model of Eq.~\eqref{lv} has been the focus of many recent investigations in theoretical ecology, yet, 
it is important to keep in mind the assumptions on which it is based. First, it neglects spatial structures present in real ecological networks. 
Second, it assumes that interspecies interactions are randomly distributed and neglects higher-order interactions involving three or more species. 
Third, and finally, a linear intrinsic growth rate $\kappa_iN_i$ is assumed, whereas some empirical data suggest a sublinear growth rate, which may have a significant impact on 
stability vs. biodiversity.~\cite{Hat24,Agu25}

From now on, our focus is on the parameters $\sigma$ and $\gamma$ defining the variability of the interspecies interactions and accordingly we fix the growth rates, the initial number of species and the average interaction at values $\kappa_i \equiv 1$, $S=157$ and $\mu=5$. 
We have checked that varying $\mu$ does not change our conclusions as long as $\mu>0$.

\begin{figure}
\includegraphics[width=.8\columnwidth]{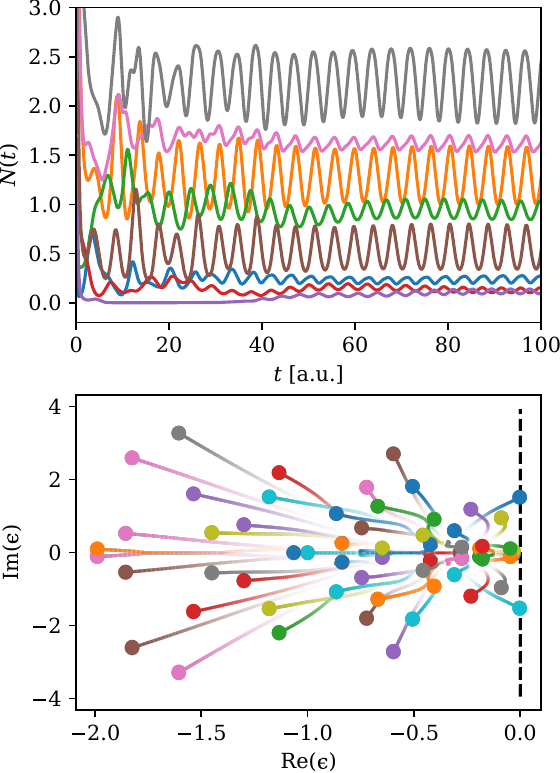}
 \caption{Top: abundances for 8 of the $N_s=61$ surviving species oscillating in a limit cycle for a realization of the model of Eq.~\eqref{lv}, with $\sigma=4.05$ and $\gamma=-0.5$. 
  Bottom: Evolution of the eigenvalues of the stability matrix ${\mathbb M}$ for $\sigma \in [0,4.02]$ (more faded colors indicate smaller values of $\sigma$). 
  Stability of the fixed point is lost when a pair of complex-conjugated eigenvalues cross the imaginary axis, resulting in a Hopf bifurcation and the emergence of the limit cycle shown in the top panel.}
   \label{fig:fig1} 
\end{figure}

The RMT approach to ecosystem dynamics dates back at least to May's seminal work.~\cite{May72} 
Fixed-points $\vec{N^*}$ of Eq.~\eqref{lv} are defined by $\dot N_i^* = 0$ and the dynamics in their vicinity is governed by the stability matrix $\mathbb{M}$
\begin{subequations} 
\begin{eqnarray}
 \delta \dot{\vec N}&=& \mathbb{M}  \, \delta \vec N \, , \\
 \label{matrel}
 \mathbb{M}_{ij} &=&-N_i^* \left( \delta_{ij}+\frac{\mu}{S} + \frac{\sigma}{\sqrt{S}} {\mathbb A}_{ij}\right) \, .
\end{eqnarray} 
\end{subequations}
The fixed-point is stable as long as the spectrum of $\mathbb{M}$ lies entirely in the left half of the complex plane. 
The average density of eigenvalues of $\mathbb{A}$  is distributed within a zero-centred ellipse in the complex plane with semi-axes $\sigma(1 + \gamma)$ [resp. $\sigma(1-\gamma)$] in the real [resp. imaginary] direction.~\cite{Som88} 
The ellipse is shifted to the left by the identity matrix [the Kronecker symbol in Eq.~\eqref{matrel}], while the constant $\mu$-term  in Eq.~\eqref{matrel} adds an outlier eigenvalue $-\mu$,~\cite{Cle23} which is irrelevant for stability when considering logistic interspecies interactions with $\mu>0$.
Assuming an homogeneous distribution of populations, $N_i^* \simeq N_0 =  {\cal O}(1)$, the fixed-point is parametrically stable for
\begin{align}\label{eq:sig-gam-bound}
 \sigma &< (1+\gamma)^{-1}\, .
\end{align}
Beyond that border, fixed-point stability may be recovered via species extinctions, $N_i^* \rightarrow 0$, except for a number $N_s<S$ of surviving species, because then, the stability matrix $\mathbb{A}^{(r)}$ is reduced by removing rows and lines from $\mathbb{A}$, corresponding to the extinct species. 
Then, the spectrum of $\mathbb{A}^{(r)}$ has an elliptic support with semi-axes $\sigma(1 \pm \gamma) \sqrt{N_s/S}$. 
Stability then bounds the number of surviving species as 
\begin{equation}
N_s/S \le [\sigma (1+\gamma)]^{-2} \, .
\end{equation}
Note that this argument says nothing about the bifurcation through which fixed-points lose their stability.

\begin{figure}
 \centering
 \hspace{-2mm}\includegraphics[width=\columnwidth]{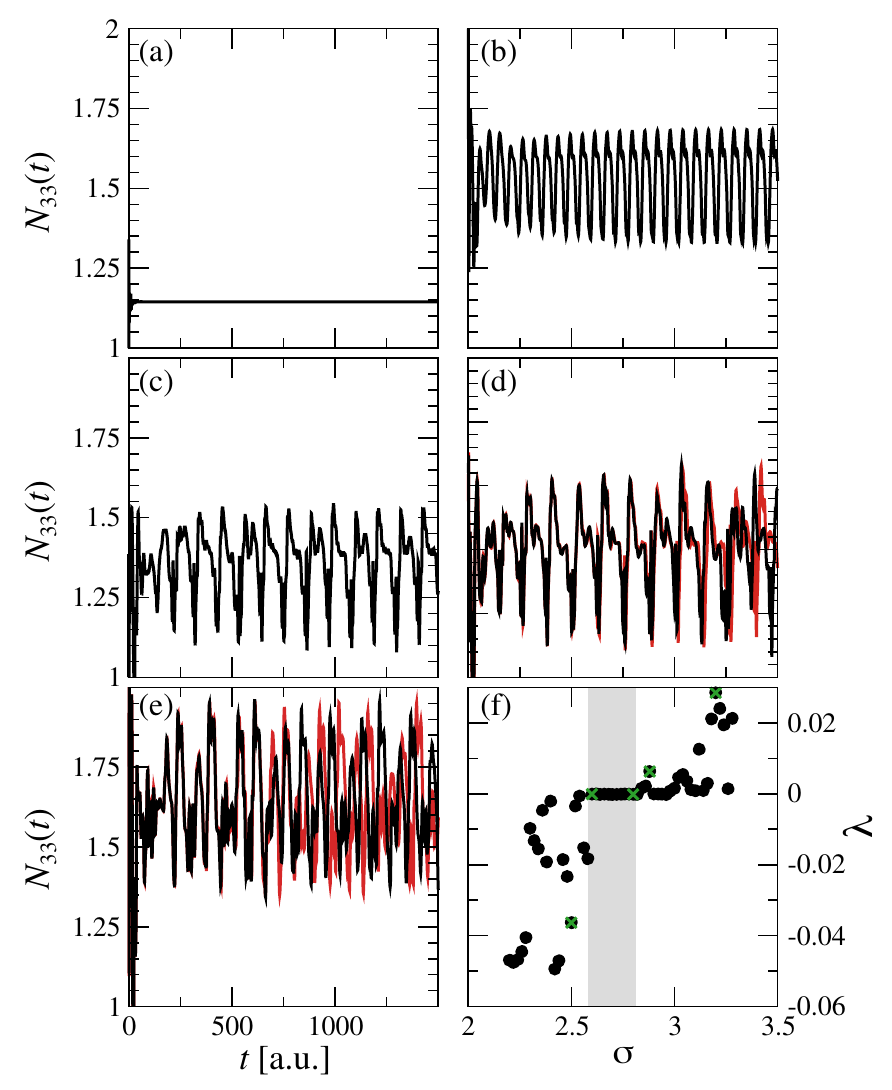}
 \caption{Emergence of a limit cycle and transition to chaos for the model of Eq.~\eqref{lv}, with $\gamma=-0.5$.
  The time-evolution of the species abundance $N_{33}$ is plotted for (a) $\sigma=2.5$, (b) $\sigma=2.6$, (c) $\sigma=2.8$, (d) $\sigma=2.85$, and (e) $\sigma=3.19$. 
  Panel (f) shows the numerically computed largest Lyapunov exponent $\lambda$, with green crosses corresponding to the cases shown in panels (a-e). 
  In panel (d) and (e), two initially nearby sets of abundances diverge from one another (black and red curves), reflecting the corresponding positive $\lambda$. 
  Different trajectories repeat similar patterns, albeit in different sequences and without periodicity. 
  Together with $\lambda > 0$, this is characteristic of the presence of a strange attractor.
  There is no such sensitivity to initial conditions in the three other cases, where $\lambda <0$ when the dynamics is attracted to a fixed point [as in panel (a)], and $\lambda=0$ (grey area) in the presence of a limit cycle [as in panels (b) and (c)].} 
 \label{fig:fig2} 
\end{figure}

\section{Results} 
For $\gamma \ne 1$ the eigenvalues of the real, asymmetric matrix $\mathbb{M}$ are either real, or come in complex-conjugated pairs. 
This opens up the possibility that the fixed-point loses its stability through a Hopf bifurcation, after which the ecosystem dynamics is attracted to a stable limit cycle. 
Such a Hopf bifurcation is illustrated in Fig.~\ref{fig:fig1}, where the top panel shows periodic oscillations in abundances of the surviving species which are related in the bottom panel to the crossing of the imaginary axis by a complex-conjugated pair of eigenvalues.
For large random matrices with $\gamma=0$ it has been shown that only ${\cal O}(\sqrt{S})$ of the $S$ eigenvalues are real,~\cite{Ede94,Ede97,Kan05} which suggests that such Hopf bifurcations are the rule rather than the exception in ecosystems with high biodiversity. 

Directly following the bifurcation, cycles exhibit sinusoidal oscillations with a single frequency, determined by the imaginary part of the pair of involved eigenvalues. 
More harmonics emerge as $\sigma$ increases further, until the cycle disappears. 
Cycle disapearance can happen because the cycle loses either its stability -- for instance through an inverse Hopf bifurcation -- or its feasibility -- for instance because one or several species reach the extinction threshold $N_c$ somewhere along the cycle. 
The dynamics of one arbitraily chosen species as $\sigma$ passes through and keeps increasing beyond a Hopf bifurcation is shown in Fig.~\ref{fig:fig2}, which illustrates a third mechanism by which a periodic cycle turns into a strange attractor. 
To quantify this transition to chaos, we numerically calculated the largest Lyapunov exponent $\lambda$.~\cite{Ben80a,Ben80b} 
As expected, $\lambda<0$ in the fixed-point regime, $\sigma \lesssim 2.59$. 
Following the Hopf bifurcation, $\lambda=0$ as long as the limit cycle remains stable [panels (b) and (c)], which corresponds to the dynamics in the direction tangential to  the cycle. 
Upon further increase of $\sigma$, we get $\lambda >0$ as one enters the chaotic regime, with a population dynamics governed by a strange attractor. 
The black and red trajectories in panels (d) and (e) illustrate the associated sensitivity to initial conditions, where population trajectories repeat similar-looking patterns, albeit at irregular time intervals and following sequences depending strongly on initial conditions. 
We stress that the abrupt fluctuations exhibited by $\lambda$ in both the fixed-point and the strange attractor regimes reflect fast dynamical changes with $\sigma$. 
Numerical error bars in panel (f) of Fig.~\ref{fig:fig2} are smaller than symbol sizes, in particular, $|\lambda| \lesssim 10^{-5}$ in the limit cycle regime $2.59 \lesssim \sigma \lesssim 2.82$ (grey area in Fig.~\ref{fig:fig2}).

\begin{figure}
 \centering
 \includegraphics[width=0.96 \columnwidth]{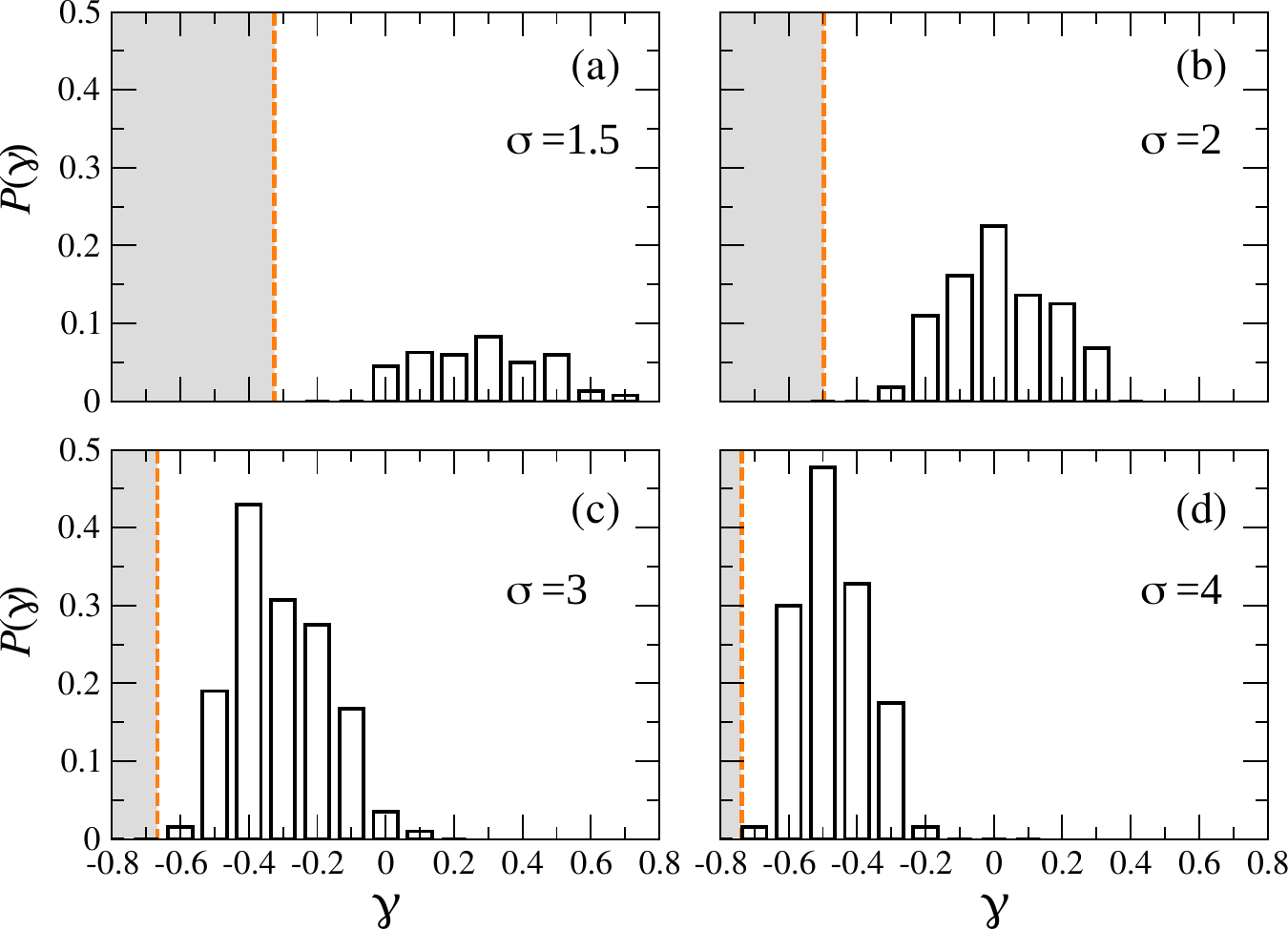}
 \caption{Probability to find the system in a state of persistent dynamics -- either a limit cycle or a state of chaotic dynamics on a strange attractor -- as a function of  $\gamma$, and for $\sigma=1.5$ [Panel(a)], 2 [(b)], 3 [(c)] and 4 [(d)]. 
  Data have been calculated over ensembles of 100 different interaction matrices, each with 5 different initial conditions.
  The grey area indicates the range of $\gamma$ where the system converges toward a unique stable fixed point, according to Eq.~\eqref{eq:sig-gam-bound}.}
 \label{fig:fig4} 
\end{figure}

The prevalence of limit cycles depends on both the variability $\sigma$ of interspecies interaction and on the off-diagonal covariance parameter $\gamma$. 
As a matter of fact, to have a limit cycle, one needs a large enough $\sigma$ to have a bifurcation in the first place, moreover, the bifurcation must also involve a pair of complex-conjugated eigenvalues of the stability matrix.
Both the value of $\sigma$ and the probability that a complex-conjugated pair of eigenvalues triggers the bifurcation depend 
on $\gamma$. 
One expects that more negative values of $\gamma$, together with the associated larger value of $\sigma$ to have a bifurcation favor the occurence of limit cycles. 
This is confirmed in Fig.~\ref{fig:fig4}, which shows the probability $P(\gamma)$ that the ecosystem has reached a state of persistent dynamics for given values of $\sigma$ and $\gamma$.
Fig.~\ref{fig:fig4} also confirms that for $\gamma<\sigma^{-1} - 1$, below the threshold given by Eq.~\eqref{eq:sig-gam-bound}, one is in the regime with a unique asymptotically stable fixed point, without persistent dynamics.
The numerical detection method employed for this statistical analysis cannot differentiate between limit cycles and chaotic motion on a strange attractor, because this would require to evaluate the largest Lyapunov exponent in each case, which is computationally time-consuming. 
However sampling those data indicate that chaotic motion sets in only at larger interactions, and constitutes a significant fraction of the data only for $\sigma=4$. 
Work to better quantify this fraction is currently underway and goes beyond the scope of this article.
Data for larger $\sigma$ are not shown as they exhibit mass extinctions, with only very few surviving species at best, except for the smallest values of $\gamma$.

The observed increased probability to find limit cycles at negative values of $\gamma$ is related to the associated increased fraction of pairs of complex-conjugated eigenvalues in the spectrum of the stability matrix. 
Neglecting inhomogeneities in the fixed-point abundances $N_i^*$, this spectrum is real for $\gamma=1$ and purely imaginary for $\gamma=-1$, and it is expectable that the fraction of real eigenvalues decreases as $\gamma$ decreases. 
To the best of our knowledge, the only mathematically rigorous result at intermediate values of $\gamma$ is that only a fraction ${\cal O}(\sqrt{S})$ of the eigenvalues are real for the Ginibre ensemble, i.e., at $\gamma=0$.~\cite{Ede94,Ede97,Kan05} 
While these observations suggest more frequent occurences of limit cycles as $\gamma$ decreases, what truly matters is whether the extreme eigenvalue with largest real part is real or complex, for a fixed realization of the interaction matrix in Eq.~\eqref{lv} with the covariance of Eq.~\eqref{randommatr}. 
Calculating how likely that is, as a function of $\gamma$, would be a formidable task and we are unaware of any rigorous result in this direction. 
We therefore resort to numerical calculations (see supplementary material, Fig.~S1).
As expected, we find that the probability of a complex extreme eigenvalue increases with decreasing $\gamma$. 
To translate this result into a probability to have a limit cycle emerge through a Hopf bifurcation, we still need to take into account that there is no bifurcation at $\gamma=-1$, since there, increasing $\sigma$ only stretches the spectrum in the imaginary direction.
We therefore expect that moderately negative values of $\gamma$ should favor the emergence of limit cycles. 
This qualitative argument is confirmed by the data shown in Fig.~\ref{fig:fig4}. 
A second numerical result shown in Fig.~\ref{fig:figS1} (see supplementary material) is that the probability that the extreme eigenvalue is complex increases with the number $S$ of species.

\begin{figure}
 \centering 
 \includegraphics[width=\columnwidth]{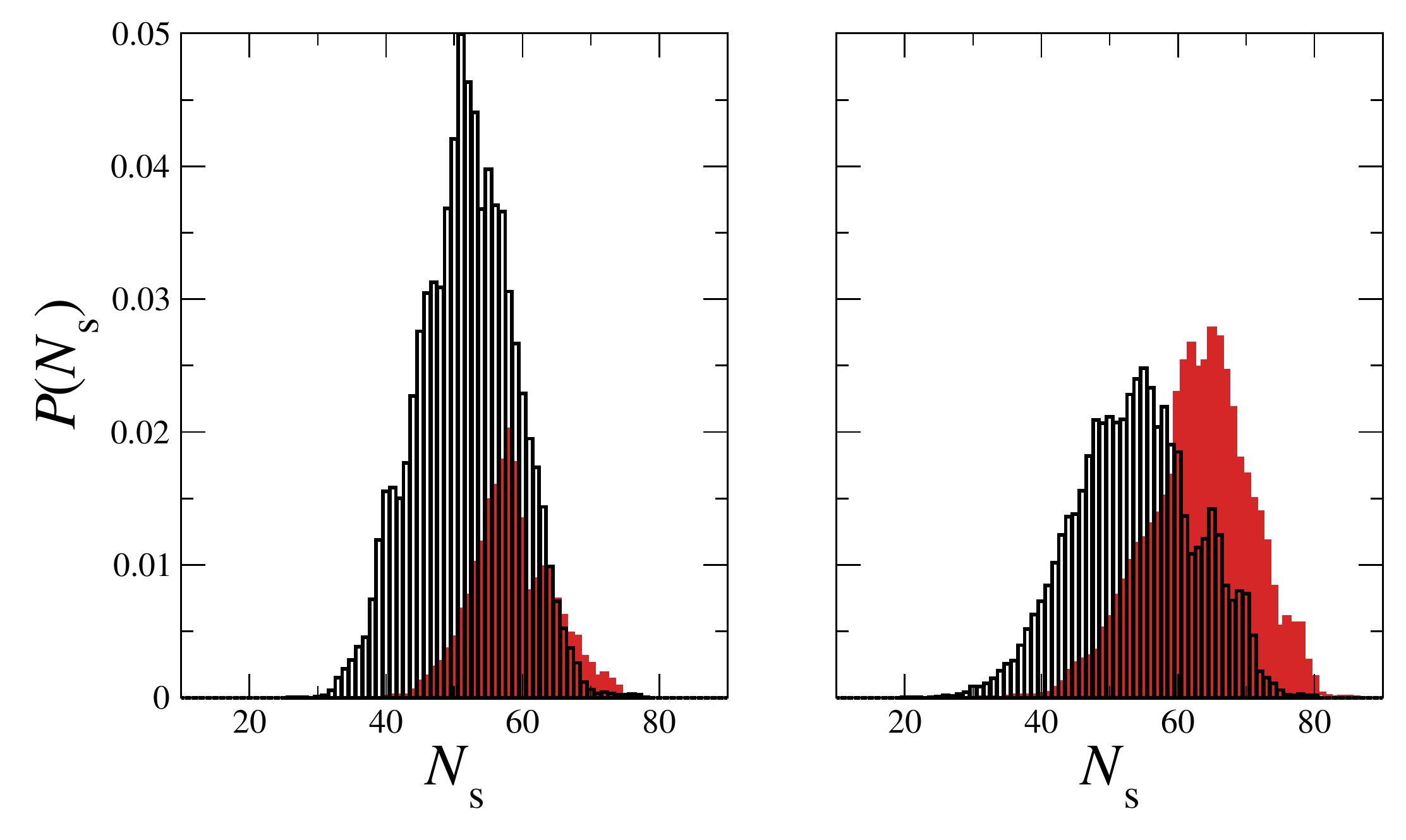}
 \caption{Distribution of the number of surviving species for the model of Eq.~\eqref{lv}, with $\sigma=2$, $\gamma=0$ (left) and $\sigma=4$, $\gamma=-0.5$ (right). 
  Black histograms correspond to fixed point solutions and red histograms to limit cycle solutions. 
  Distributions are calculated over 400 realizations of the interaction matrix ${\mathbb A}_{ij}$, each with 25 different initial abundances. 
  In all considered cases, the system converges to either a fixed point or a limit cycle.
  Distributions correspond to  22.5\% (left panel) and 47.6 \% of cycles (right panel).}
 \label{fig:fig3}
\end{figure}

We translate these results into the language of theoretical ecology. 
Noting that $\gamma$ interpolates between ecosystems consisting purely of pairs of either mutualistic or competitive species for $\gamma=1$, to ecosystems with only predator-prey pairs of species for $\gamma=-1$~\cite{gamma}, persistent oscillating behavior in population abundances are expected to be prevailing in large, sufficiently interacting ecosystems with a majority of predator-prey pairs. 
We note that the occurence of Hopf bifurcations is of importance for theoretical ecology. 
It extends the stability of population coexistence beyond the loss of fixed-point stability, without the need for species extinction. 
The resulting increasing biodiversity is illustrated in Fig.~\ref{fig:fig3}. 
It is obvious that the average number of coexisting species at fixed $\sigma$ is higher for ecosystems equilibrating to limit cycles ($\langle N_s\rangle =58.4$ and $63$) than to fixed-points ($\langle N_s\rangle =51.8$ and $53.7$). 

\section{Conclusions}
It has long been known that multi-species ecosystems described by sets of coupled ordinary differential equations of the type given in Eq.~\eqref{lv} may exhibit any dynamical behavior.~\cite{Sma76} 
Here we have emphasized a route joining the previously observed phases governed by attractive fixed-points to those exhibiting persistent dynamics. 
Because species in ecosystems and food webs interact with one another in a necessarily asymmetric way, fixed-point instabilities may occur via a Hopf bifurcation.
The resulting limit cycles preserve biodiversity, since at least for some range of interaction variability, their stability does not necessitate species extinctions. 

Models of theoretical ecology are not expected to precisely reflect ecosystem behaviors. 
Instead, they may shed qualitative light on different, observed behaviors of  real or lab-controlled ecosystems at a statistical level. 
In that respect, our work emphasizes the existence of various dynamical behaviors, depending on the two parameters $\sigma$ and $\gamma$ governing interspecies interactions. 
Recent analysis of population time series found evidence of chaotic behavior in at least 30\% of the studied populations.~\cite{Rog22}
Fig.~\ref{fig:fig2} further emphasizes parametrically sizable regimes with small Lyapunov exponents, which has also been observed.~\cite{Rog22,Ell95} 
Other regimes with strong sensitivity to even minor parametric changes [e.g., at $\sigma \gtrsim 3$ in Fig.~\ref{fig:fig2}(f)] are characteristic of systems close to criticality as discussed in Refs.~\onlinecite{Sol02,Mor11}.
Since criticality slows down the dynamics, the behavior of ecosystems in this latter regime is transient by nature and not governed by any long-time asymptotic. 
That ecosystems are governed by transient behaviors has been postulated in Refs.~\onlinecite{Hu22,Has94,Schr03,Mor16,Has18}. 
The emergence of Hopf bifurcations at negative values of $\gamma$ finally explains recent results where ecosystems with dominating predator-prey interactions display oscillatory behaviors in their population dynamics.~\cite{Mam22}

The present work illustrates that all the different, observed or theoretically postulated behaviors naturally emerge from a single, unified model, without the need for exogeneous intervention. 
The task at hand now is obviously to try and determine, at least qualitatively, model parameters corresponding to specifically observed ecosystems. 
Our work has simplified that task in that it identified the two key model parameters $\sigma$ and $\gamma$ driving transitions between different dynamical behaviors. 

There are several important direction in which our work should be extended. 
Among them, we mention first, that we are currently investigating the frequency of occurence of strange attractors vs. restabilization after species extinction at large $\sigma$. 
Second, further investigations should extend our results to interaction matrices reflecting more realistic topologies of known ecological networks.~\cite{Fri17} 
Third, investigating species distributions may give precious information on the conditions under which biodiversity and rarity may coexist.~\cite{Sol02} 
Fourth, ecosystem parameters are modified by climate changes.~\cite{Ral10,Eng11,Par19}
Investigating changes in ecosystem functioning under climate-induced changes in trophic interactions is of paramount interest. 
It would evidently have a strong influence on ecosystem functioning in the critical and chaotic regimes with strong parameter sensitivity.
There are certainly many other interesting extensions.

\section*{Supplementary Material}
Section I of the supplementary material provides a numerical estimate of the probability that the extreme eigenvalue of the matrix in Eq.~\eqref{randommatr} is complex.
In Sec. II of the supplementary material, we numerically illustrate various dynamical behaviors that are observed in the system of Eq.~\eqref{lv}.

\section*{Acknowledgments} 
We thank Christian Mazza and Xavier Richard for introducing us to theoretical ecology and population dynamics. 
RD was supported by the Swiss National Science Foundation, under grant nr. 200021\_215336.

\section*{Author declaration}
\subsection*{Conflict of interest}
The authors have no conflict to disclose.

\subsection*{Author contributions}
{\bf Robin Delabays:}
Conceptualization (support);
Formal analysis (support);
Investigation (support);
Methodology (support);
Visualization (equal);
Writing -- original draft (support);
Writing -- review \& editing (equal).

{\bf Philippe Jacquod:}
Conceptualization (lead);
Formal analysis (lead);
Investigation (lead);
Methodology (lead);
Visualization (equal);
Writing -- original draft (lead);
Writing -- review \& editing (equal).

\section*{Data availability statement}
The data that support the findings of this study are available within the article.


\begin{thebibliography}{57}%
\makeatletter
\providecommand \@ifxundefined [1]{%
 \@ifx{#1\undefined}
}%
\providecommand \@ifnum [1]{%
 \ifnum #1\expandafter \@firstoftwo
 \else \expandafter \@secondoftwo
 \fi
}%
\providecommand \@ifx [1]{%
 \ifx #1\expandafter \@firstoftwo
 \else \expandafter \@secondoftwo
 \fi
}%
\providecommand \natexlab [1]{#1}%
\providecommand \enquote  [1]{``#1''}%
\providecommand \bibnamefont  [1]{#1}%
\providecommand \bibfnamefont [1]{#1}%
\providecommand \citenamefont [1]{#1}%
\providecommand \href@noop [0]{\@secondoftwo}%
\providecommand \href [0]{\begingroup \@sanitize@url \@href}%
\providecommand \@href[1]{\@@startlink{#1}\@@href}%
\providecommand \@@href[1]{\endgroup#1\@@endlink}%
\providecommand \@sanitize@url [0]{\catcode `\\12\catcode `\$12\catcode
  `\&12\catcode `\#12\catcode `\^12\catcode `\_12\catcode `\%12\relax}%
\providecommand \@@startlink[1]{}%
\providecommand \@@endlink[0]{}%
\providecommand \url  [0]{\begingroup\@sanitize@url \@url }%
\providecommand \@url [1]{\endgroup\@href {#1}{\urlprefix }}%
\providecommand \urlprefix  [0]{URL }%
\providecommand \Eprint [0]{\href }%
\providecommand \doibase [0]{http://dx.doi.org/}%
\providecommand \selectlanguage [0]{\@gobble}%
\providecommand \bibinfo  [0]{\@secondoftwo}%
\providecommand \bibfield  [0]{\@secondoftwo}%
\providecommand \translation [1]{[#1]}%
\providecommand \BibitemOpen [0]{}%
\providecommand \bibitemStop [0]{}%
\providecommand \bibitemNoStop [0]{.\EOS\space}%
\providecommand \EOS [0]{\spacefactor3000\relax}%
\providecommand \BibitemShut  [1]{\csname bibitem#1\endcsname}%
\let\auto@bib@innerbib\@empty
\bibitem [{\citenamefont {{{May, Robert and McLean, Angela R}}}(2007)}]{May07}%
  \BibitemOpen
  \bibinfo {editor} {\bibnamefont {{{May, Robert and McLean, Angela R}}}},\
  ed.,\ \href {\doibase 10.1093/oso/9780199209989.001.0001} {\emph {\bibinfo
  {title} {Theoretical {{Ecology}}: {{Principles}} and {{Applications}}}}}\
  (\bibinfo  {publisher} {Oxford University Press},\ \bibinfo {year}
  {2007})\BibitemShut {NoStop}%
\bibitem [{\citenamefont {Ranta}, \citenamefont {Kaitala},\ and\ \citenamefont
  {Lundberg}(1998)}]{Ran98}%
  \BibitemOpen
  \bibfield  {author} {\bibinfo {author} {\bibfnamefont {E.}~\bibnamefont
  {Ranta}}, \bibinfo {author} {\bibfnamefont {V.}~\bibnamefont {Kaitala}}, \
  and\ \bibinfo {author} {\bibfnamefont {P.}~\bibnamefont {Lundberg}},\
  }\bibfield  {title} {\enquote {\bibinfo {title} {Population {{Variability}}
  in {{Space}} and {{Time}}: {{The Dynamics}} of {{Synchronous Population
  Fluctuations}}},}\ }\href {\doibase 10.2307/3546852} {\bibfield  {journal}
  {\bibinfo  {journal} {Oikos}\ }\textbf {\bibinfo {volume} {83}},\ \bibinfo
  {pages} {376--382} (\bibinfo {year} {1998})},\ \Eprint
  {http://arxiv.org/abs/3546852} {3546852} \BibitemShut {NoStop}%
\bibitem [{\citenamefont {Lundberg}\ \emph {et~al.}(2000)\citenamefont
  {Lundberg}, \citenamefont {Ranta}, \citenamefont {Ripa},\ and\ \citenamefont
  {Kaitala}}]{Lun00}%
  \BibitemOpen
  \bibfield  {author} {\bibinfo {author} {\bibfnamefont {P.}~\bibnamefont
  {Lundberg}}, \bibinfo {author} {\bibfnamefont {E.}~\bibnamefont {Ranta}},
  \bibinfo {author} {\bibfnamefont {J.}~\bibnamefont {Ripa}}, \ and\ \bibinfo
  {author} {\bibfnamefont {V.}~\bibnamefont {Kaitala}},\ }\bibfield  {title}
  {\enquote {\bibinfo {title} {Population variability in space and time},}\
  }\href {\doibase 10.1016/S0169-5347(00)01981-9} {\bibfield  {journal}
  {\bibinfo  {journal} {Trends in Ecology \& Evolution}\ }\textbf {\bibinfo
  {volume} {15}},\ \bibinfo {pages} {460--464} (\bibinfo {year}
  {2000})}\BibitemShut {NoStop}%
\bibitem [{\citenamefont {Inchausti}\ and\ \citenamefont
  {Halley}(2003)}]{Inc03}%
  \BibitemOpen
  \bibfield  {author} {\bibinfo {author} {\bibfnamefont {P.}~\bibnamefont
  {Inchausti}}\ and\ \bibinfo {author} {\bibfnamefont {J.}~\bibnamefont
  {Halley}},\ }\bibfield  {title} {\enquote {\bibinfo {title} {On the relation
  between temporal variability and persistence time in animal populations},}\
  }\href {\doibase 10.1046/j.1365-2656.2003.00767.x} {\bibfield  {journal}
  {\bibinfo  {journal} {Journal of Animal Ecology}\ }\textbf {\bibinfo {volume}
  {72}},\ \bibinfo {pages} {899--908} (\bibinfo {year} {2003})}\BibitemShut
  {NoStop}%
\bibitem [{\citenamefont {Vandermeer}(2006)}]{Van06}%
  \BibitemOpen
  \bibfield  {author} {\bibinfo {author} {\bibfnamefont {J.}~\bibnamefont
  {Vandermeer}},\ }\bibfield  {title} {\enquote {\bibinfo {title} {Oscillating
  {{Populations}} and {{Biodiversity Maintenance}}},}\ }\href {\doibase
  10.1641/0006-3568(2006)56[967:OPABM]2.0.CO;2} {\bibfield  {journal} {\bibinfo
   {journal} {BioScience}\ }\textbf {\bibinfo {volume} {56}},\ \bibinfo {pages}
  {967--975} (\bibinfo {year} {2006})}\BibitemShut {NoStop}%
\bibitem [{\citenamefont {Hastings}\ \emph {et~al.}(1993)\citenamefont
  {Hastings}, \citenamefont {Hom}, \citenamefont {Ellner}, \citenamefont
  {Turchin},\ and\ \citenamefont {Godfray}}]{Has93}%
  \BibitemOpen
  \bibfield  {author} {\bibinfo {author} {\bibfnamefont {A.}~\bibnamefont
  {Hastings}}, \bibinfo {author} {\bibfnamefont {C.~L.}\ \bibnamefont {Hom}},
  \bibinfo {author} {\bibfnamefont {S.}~\bibnamefont {Ellner}}, \bibinfo
  {author} {\bibfnamefont {P.}~\bibnamefont {Turchin}}, \ and\ \bibinfo
  {author} {\bibfnamefont {H.~C.~J.}\ \bibnamefont {Godfray}},\ }\bibfield
  {title} {\enquote {\bibinfo {title} {Chaos in {{Ecology}}: {{Is Mother
  Nature}} a {{Strange Attractor}}?*},}\ }\href {\doibase
  10.1146/annurev.es.24.110193.000245} {\bibfield  {journal} {\bibinfo
  {journal} {Annual Review of Ecology, Evolution, and Systematics}\ }\textbf
  {\bibinfo {volume} {24}},\ \bibinfo {pages} {1--33} (\bibinfo {year}
  {1993})}\BibitemShut {NoStop}%
\bibitem [{\citenamefont {Ellner}\ and\ \citenamefont {Turchin}(1995)}]{Ell95}%
  \BibitemOpen
  \bibfield  {author} {\bibinfo {author} {\bibfnamefont {S.}~\bibnamefont
  {Ellner}}\ and\ \bibinfo {author} {\bibfnamefont {P.}~\bibnamefont
  {Turchin}},\ }\bibfield  {title} {\enquote {\bibinfo {title} {Chaos in a
  {{Noisy World}}: {{New Methods}} and {{Evidence}} from {{Time-Series
  Analysis}}},}\ }\href {\doibase 10.1086/285744} {\bibfield  {journal}
  {\bibinfo  {journal} {The American Naturalist}\ }\textbf {\bibinfo {volume}
  {145}},\ \bibinfo {pages} {343--375} (\bibinfo {year} {1995})}\BibitemShut
  {NoStop}%
\bibitem [{\citenamefont {Hunter}\ and\ \citenamefont {Price}(1998)}]{Hun98}%
  \BibitemOpen
  \bibfield  {author} {\bibinfo {author} {\bibfnamefont {M.~D.}\ \bibnamefont
  {Hunter}}\ and\ \bibinfo {author} {\bibfnamefont {P.~W.}\ \bibnamefont
  {Price}},\ }\bibfield  {title} {\enquote {\bibinfo {title} {Cycles in insect
  populations: Delayed density dependence or exogenous driving variables?}}\
  }\href {\doibase 10.1046/j.1365-2311.1998.00123.x} {\bibfield  {journal}
  {\bibinfo  {journal} {Ecological Entomology}\ }\textbf {\bibinfo {volume}
  {23}},\ \bibinfo {pages} {216--222} (\bibinfo {year} {1998})}\BibitemShut
  {NoStop}%
\bibitem [{\citenamefont {Huisman}\ and\ \citenamefont
  {Weissing}(1999)}]{Hui99}%
  \BibitemOpen
  \bibfield  {author} {\bibinfo {author} {\bibfnamefont {J.}~\bibnamefont
  {Huisman}}\ and\ \bibinfo {author} {\bibfnamefont {F.~J.}\ \bibnamefont
  {Weissing}},\ }\bibfield  {title} {\enquote {\bibinfo {title} {Biodiversity
  of plankton by species oscillations and chaos},}\ }\href {\doibase
  10.1038/46540} {\bibfield  {journal} {\bibinfo  {journal} {Nature}\ }\textbf
  {\bibinfo {volume} {402}},\ \bibinfo {pages} {407--410} (\bibinfo {year}
  {1999})}\BibitemShut {NoStop}%
\bibitem [{\citenamefont {Beninc{\`a}}\ \emph {et~al.}(2008)\citenamefont
  {Beninc{\`a}}, \citenamefont {Huisman}, \citenamefont {Heerkloss},
  \citenamefont {J{\"o}hnk}, \citenamefont {Branco}, \citenamefont {Van~Nes},
  \citenamefont {Scheffer},\ and\ \citenamefont {Ellner}}]{Ben08}%
  \BibitemOpen
  \bibfield  {author} {\bibinfo {author} {\bibfnamefont {E.}~\bibnamefont
  {Beninc{\`a}}}, \bibinfo {author} {\bibfnamefont {J.}~\bibnamefont
  {Huisman}}, \bibinfo {author} {\bibfnamefont {R.}~\bibnamefont {Heerkloss}},
  \bibinfo {author} {\bibfnamefont {K.~D.}\ \bibnamefont {J{\"o}hnk}}, \bibinfo
  {author} {\bibfnamefont {P.}~\bibnamefont {Branco}}, \bibinfo {author}
  {\bibfnamefont {E.~H.}\ \bibnamefont {Van~Nes}}, \bibinfo {author}
  {\bibfnamefont {M.}~\bibnamefont {Scheffer}}, \ and\ \bibinfo {author}
  {\bibfnamefont {S.~P.}\ \bibnamefont {Ellner}},\ }\bibfield  {title}
  {\enquote {\bibinfo {title} {Chaos in a long-term experiment with a plankton
  community},}\ }\href {\doibase 10.1038/nature06512} {\bibfield  {journal}
  {\bibinfo  {journal} {Nature}\ }\textbf {\bibinfo {volume} {451}},\ \bibinfo
  {pages} {822--825} (\bibinfo {year} {2008})}\BibitemShut {NoStop}%
\bibitem [{\citenamefont {Rogers}, \citenamefont {Johnson},\ and\ \citenamefont
  {Munch}(2022)}]{Rog22}%
  \BibitemOpen
  \bibfield  {author} {\bibinfo {author} {\bibfnamefont {T.~L.}\ \bibnamefont
  {Rogers}}, \bibinfo {author} {\bibfnamefont {B.~J.}\ \bibnamefont {Johnson}},
  \ and\ \bibinfo {author} {\bibfnamefont {S.~B.}\ \bibnamefont {Munch}},\
  }\bibfield  {title} {\enquote {\bibinfo {title} {Chaos is not rare in natural
  ecosystems},}\ }\href {\doibase 10.1038/s41559-022-01787-y} {\bibfield
  {journal} {\bibinfo  {journal} {Nature Ecology \& Evolution}\ }\textbf
  {\bibinfo {volume} {6}},\ \bibinfo {pages} {1105--1111} (\bibinfo {year}
  {2022})}\BibitemShut {NoStop}%
\bibitem [{\citenamefont {Prendergast}\ \emph {et~al.}(2010)\citenamefont
  {Prendergast}, \citenamefont {{Bazeley-White}}, \citenamefont {Smith},
  \citenamefont {Lawton}, \citenamefont {Inchausti}, \citenamefont {Kidd},\
  and\ \citenamefont {Knight}}]{Pre10}%
  \BibitemOpen
  \bibfield  {author} {\bibinfo {author} {\bibfnamefont {J.}~\bibnamefont
  {Prendergast}}, \bibinfo {author} {\bibfnamefont {E.}~\bibnamefont
  {{Bazeley-White}}}, \bibinfo {author} {\bibfnamefont {O.}~\bibnamefont
  {Smith}}, \bibinfo {author} {\bibfnamefont {J.}~\bibnamefont {Lawton}},
  \bibinfo {author} {\bibfnamefont {P.}~\bibnamefont {Inchausti}}, \bibinfo
  {author} {\bibfnamefont {D.}~\bibnamefont {Kidd}}, \ and\ \bibinfo {author}
  {\bibfnamefont {S.}~\bibnamefont {Knight}},\ }\bibfield  {title} {\enquote
  {\bibinfo {title} {The {{Global Population Dynamics Database}}},}\ }\href
  {\doibase 10.5063/F1BZ63Z8} {\  (\bibinfo {year} {2010}),\
  10.5063/F1BZ63Z8}\BibitemShut {NoStop}%
\bibitem [{\citenamefont {Pearce}, \citenamefont {Agarwala},\ and\
  \citenamefont {Fisher}(2020)}]{Pea20}%
  \BibitemOpen
  \bibfield  {author} {\bibinfo {author} {\bibfnamefont {M.~T.}\ \bibnamefont
  {Pearce}}, \bibinfo {author} {\bibfnamefont {A.}~\bibnamefont {Agarwala}}, \
  and\ \bibinfo {author} {\bibfnamefont {D.~S.}\ \bibnamefont {Fisher}},\
  }\bibfield  {title} {\enquote {\bibinfo {title} {Stabilization of extensive
  fine-scale diversity by ecologically driven spatiotemporal chaos},}\ }\href
  {\doibase 10.1073/pnas.1915313117} {\bibfield  {journal} {\bibinfo  {journal}
  {Proceedings of the National Academy of Sciences}\ }\textbf {\bibinfo
  {volume} {117}},\ \bibinfo {pages} {14572--14583} (\bibinfo {year}
  {2020})}\BibitemShut {NoStop}%
\bibitem [{\citenamefont {Scheffer}\ \emph {et~al.}(2003)\citenamefont
  {Scheffer}, \citenamefont {Rinaldi}, \citenamefont {Huisman},\ and\
  \citenamefont {Weissing}}]{Sch03}%
  \BibitemOpen
  \bibfield  {author} {\bibinfo {author} {\bibfnamefont {M.}~\bibnamefont
  {Scheffer}}, \bibinfo {author} {\bibfnamefont {S.}~\bibnamefont {Rinaldi}},
  \bibinfo {author} {\bibfnamefont {J.}~\bibnamefont {Huisman}}, \ and\
  \bibinfo {author} {\bibfnamefont {F.~J.}\ \bibnamefont {Weissing}},\
  }\bibfield  {title} {\enquote {\bibinfo {title} {Why plankton communities
  have no equilibrium: Solutions to the paradox},}\ }\href {\doibase
  10.1023/A:1024404804748} {\bibfield  {journal} {\bibinfo  {journal}
  {Hydrobiologia}\ }\textbf {\bibinfo {volume} {491}},\ \bibinfo {pages}
  {9--18} (\bibinfo {year} {2003})}\BibitemShut {NoStop}%
\bibitem [{\citenamefont {May}(1972)}]{May72}%
  \BibitemOpen
  \bibfield  {author} {\bibinfo {author} {\bibfnamefont {R.~M.}\ \bibnamefont
  {May}},\ }\bibfield  {title} {\enquote {\bibinfo {title} {Will a {{Large
  Complex System}} be {{Stable}}?}}\ }\href {\doibase 10.1038/238413a0}
  {\bibfield  {journal} {\bibinfo  {journal} {Nature}\ }\textbf {\bibinfo
  {volume} {238}},\ \bibinfo {pages} {413--414} (\bibinfo {year}
  {1972})}\BibitemShut {NoStop}%
\bibitem [{\citenamefont {Roberts}(1974)}]{Rob74}%
  \BibitemOpen
  \bibfield  {author} {\bibinfo {author} {\bibfnamefont {A.}~\bibnamefont
  {Roberts}},\ }\bibfield  {title} {\enquote {\bibinfo {title} {The stability
  of a feasible random ecosystem},}\ }\href {\doibase 10.1038/251607a0}
  {\bibfield  {journal} {\bibinfo  {journal} {Nature}\ }\textbf {\bibinfo
  {volume} {251}},\ \bibinfo {pages} {607--608} (\bibinfo {year}
  {1974})}\BibitemShut {NoStop}%
\bibitem [{\citenamefont {Rieger}(1989)}]{Rie89}%
  \BibitemOpen
  \bibfield  {author} {\bibinfo {author} {\bibfnamefont {H.}~\bibnamefont
  {Rieger}},\ }\bibfield  {title} {\enquote {\bibinfo {title} {Solvable model
  of a complex ecosystem with randomly interacting species},}\ }\href {\doibase
  10.1088/0305-4470/22/17/011} {\bibfield  {journal} {\bibinfo  {journal}
  {Journal of Physics A: Mathematical and General}\ }\textbf {\bibinfo {volume}
  {22}},\ \bibinfo {pages} {3447} (\bibinfo {year} {1989})}\BibitemShut
  {NoStop}%
\bibitem [{\citenamefont {Allesina}\ and\ \citenamefont {Tang}(2012)}]{All12}%
  \BibitemOpen
  \bibfield  {author} {\bibinfo {author} {\bibfnamefont {S.}~\bibnamefont
  {Allesina}}\ and\ \bibinfo {author} {\bibfnamefont {S.}~\bibnamefont
  {Tang}},\ }\bibfield  {title} {\enquote {\bibinfo {title} {Stability criteria
  for complex ecosystems},}\ }\href {\doibase 10.1038/nature10832} {\bibfield
  {journal} {\bibinfo  {journal} {Nature}\ }\textbf {\bibinfo {volume} {483}},\
  \bibinfo {pages} {205--208} (\bibinfo {year} {2012})}\BibitemShut {NoStop}%
\bibitem [{\citenamefont {Tang}, \citenamefont {Pawar},\ and\ \citenamefont
  {Allesina}(2014)}]{Tan14}%
  \BibitemOpen
  \bibfield  {author} {\bibinfo {author} {\bibfnamefont {S.}~\bibnamefont
  {Tang}}, \bibinfo {author} {\bibfnamefont {S.}~\bibnamefont {Pawar}}, \ and\
  \bibinfo {author} {\bibfnamefont {S.}~\bibnamefont {Allesina}},\ }\bibfield
  {title} {\enquote {\bibinfo {title} {Correlation between interaction
  strengths drives stability in large ecological networks},}\ }\href {\doibase
  10.1111/ele.12312} {\bibfield  {journal} {\bibinfo  {journal} {Ecology
  Letters}\ }\textbf {\bibinfo {volume} {17}},\ \bibinfo {pages} {1094--1100}
  (\bibinfo {year} {2014})}\BibitemShut {NoStop}%
\bibitem [{\citenamefont {Fyodorov}\ and\ \citenamefont
  {Khoruzhenko}(2016)}]{Fyo16}%
  \BibitemOpen
  \bibfield  {author} {\bibinfo {author} {\bibfnamefont {Y.~V.}\ \bibnamefont
  {Fyodorov}}\ and\ \bibinfo {author} {\bibfnamefont {B.~A.}\ \bibnamefont
  {Khoruzhenko}},\ }\bibfield  {title} {\enquote {\bibinfo {title} {Nonlinear
  analogue of the {{May}}-{{Wigner}} instability transition},}\ }\href
  {\doibase 10.1073/pnas.1601136113} {\bibfield  {journal} {\bibinfo  {journal}
  {Proceedings of the National Academy of Sciences}\ }\textbf {\bibinfo
  {volume} {113}},\ \bibinfo {pages} {6827--6832} (\bibinfo {year}
  {2016})}\BibitemShut {NoStop}%
\bibitem [{\citenamefont {Clenet}, \citenamefont {Massol},\ and\ \citenamefont
  {Najim}(2023)}]{Cle23}%
  \BibitemOpen
  \bibfield  {author} {\bibinfo {author} {\bibfnamefont {M.}~\bibnamefont
  {Clenet}}, \bibinfo {author} {\bibfnamefont {F.}~\bibnamefont {Massol}}, \
  and\ \bibinfo {author} {\bibfnamefont {J.}~\bibnamefont {Najim}},\ }\bibfield
   {title} {\enquote {\bibinfo {title} {Equilibrium and surviving species in a
  large {{Lotka}}--{{Volterra}} system of differential equations},}\ }\href
  {\doibase 10.1007/s00285-023-01939-z} {\bibfield  {journal} {\bibinfo
  {journal} {Journal of Mathematical Biology}\ }\textbf {\bibinfo {volume}
  {87}},\ \bibinfo {pages} {13} (\bibinfo {year} {2023})}\BibitemShut {NoStop}%
\bibitem [{\citenamefont {Ros}\ \emph {et~al.}(2023)\citenamefont {Ros},
  \citenamefont {Roy}, \citenamefont {Biroli}, \citenamefont {Bunin},\ and\
  \citenamefont {Turner}}]{Ros23}%
  \BibitemOpen
  \bibfield  {author} {\bibinfo {author} {\bibfnamefont {V.}~\bibnamefont
  {Ros}}, \bibinfo {author} {\bibfnamefont {F.}~\bibnamefont {Roy}}, \bibinfo
  {author} {\bibfnamefont {G.}~\bibnamefont {Biroli}}, \bibinfo {author}
  {\bibfnamefont {G.}~\bibnamefont {Bunin}}, \ and\ \bibinfo {author}
  {\bibfnamefont {A.~M.}\ \bibnamefont {Turner}},\ }\bibfield  {title}
  {\enquote {\bibinfo {title} {Generalized {Lotka}-{Volterra} {Equations} with
  {Random}, {Nonreciprocal} {Interactions}: {The} {Typical} {Number} of
  {Equilibria}},}\ }\href {\doibase 10.1103/PhysRevLett.130.257401} {\bibfield
  {journal} {\bibinfo  {journal} {Phys. Rev. Lett.}\ }\textbf {\bibinfo
  {volume} {130}},\ \bibinfo {pages} {257401} (\bibinfo {year}
  {2023})}\BibitemShut {NoStop}%
\bibitem [{\citenamefont {Gilpin}(1975)}]{Gil75}%
  \BibitemOpen
  \bibfield  {author} {\bibinfo {author} {\bibfnamefont {M.~E.}\ \bibnamefont
  {Gilpin}},\ }\bibfield  {title} {\enquote {\bibinfo {title} {Limit {{Cycles}}
  in {{Competition Communities}}},}\ }\href {\doibase 10.1086/282973}
  {\bibfield  {journal} {\bibinfo  {journal} {The American Naturalist}\
  }\textbf {\bibinfo {volume} {109}},\ \bibinfo {pages} {51--60} (\bibinfo
  {year} {1975})}\BibitemShut {NoStop}%
\bibitem [{\citenamefont {Coste}, \citenamefont {Peyraud},\ and\ \citenamefont
  {Coullet}(1979)}]{Cos79}%
  \BibitemOpen
  \bibfield  {author} {\bibinfo {author} {\bibfnamefont {J.}~\bibnamefont
  {Coste}}, \bibinfo {author} {\bibfnamefont {J.}~\bibnamefont {Peyraud}}, \
  and\ \bibinfo {author} {\bibfnamefont {P.}~\bibnamefont {Coullet}},\
  }\bibfield  {title} {\enquote {\bibinfo {title} {Asymptotic {{Behaviors}} in
  the {{Dynamics}} of {{Competing Species}}},}\ }\href {\doibase
  10.1137/0136039} {\bibfield  {journal} {\bibinfo  {journal} {SIAM Journal on
  Applied Mathematics}\ }\textbf {\bibinfo {volume} {36}},\ \bibinfo {pages}
  {516--543} (\bibinfo {year} {1979})}\BibitemShut {NoStop}%
\bibitem [{\citenamefont {Gardini}, \citenamefont {Lupini},\ and\ \citenamefont
  {Messia}(1989)}]{Gar89}%
  \BibitemOpen
  \bibfield  {author} {\bibinfo {author} {\bibfnamefont {L.}~\bibnamefont
  {Gardini}}, \bibinfo {author} {\bibfnamefont {R.}~\bibnamefont {Lupini}}, \
  and\ \bibinfo {author} {\bibfnamefont {M.~G.}\ \bibnamefont {Messia}},\
  }\bibfield  {title} {\enquote {\bibinfo {title} {Hopf bifurcation and
  transition to chaos in {{Lotka-Volterra}} equation},}\ }\href {\doibase
  10.1007/BF00275811} {\bibfield  {journal} {\bibinfo  {journal} {Journal of
  Mathematical Biology}\ }\textbf {\bibinfo {volume} {27}},\ \bibinfo {pages}
  {259--272} (\bibinfo {year} {1989})}\BibitemShut {NoStop}%
\bibitem [{\citenamefont {Fussmann}\ \emph {et~al.}(2000)\citenamefont
  {Fussmann}, \citenamefont {Ellner}, \citenamefont {Shertzer},\ and\
  \citenamefont {Hairston~Jr.}}]{Fus00}%
  \BibitemOpen
  \bibfield  {author} {\bibinfo {author} {\bibfnamefont {G.~F.}\ \bibnamefont
  {Fussmann}}, \bibinfo {author} {\bibfnamefont {S.~P.}\ \bibnamefont
  {Ellner}}, \bibinfo {author} {\bibfnamefont {K.~W.}\ \bibnamefont
  {Shertzer}}, \ and\ \bibinfo {author} {\bibfnamefont {N.~G.}\ \bibnamefont
  {Hairston~Jr.}},\ }\bibfield  {title} {\enquote {\bibinfo {title} {Crossing
  the {{Hopf Bifurcation}} in a {{Live Predator-Prey System}}},}\ }\href
  {\doibase 10.1126/science.290.5495.1358} {\bibfield  {journal} {\bibinfo
  {journal} {Science}\ }\textbf {\bibinfo {volume} {290}},\ \bibinfo {pages}
  {1358--1360} (\bibinfo {year} {2000})}\BibitemShut {NoStop}%
\bibitem [{\citenamefont {Mcgehee}\ \emph {et~al.}(2008)\citenamefont
  {Mcgehee}, \citenamefont {Schutt}, \citenamefont {Vasquez},\ and\
  \citenamefont {Peacock-L\'opez}}]{Mcg08}%
  \BibitemOpen
  \bibfield  {author} {\bibinfo {author} {\bibfnamefont {E.~A.}\ \bibnamefont
  {Mcgehee}}, \bibinfo {author} {\bibfnamefont {N.}~\bibnamefont {Schutt}},
  \bibinfo {author} {\bibfnamefont {D.~A.}\ \bibnamefont {Vasquez}}, \ and\
  \bibinfo {author} {\bibfnamefont {E.}~\bibnamefont {Peacock-L\'opez}},\
  }\bibfield  {title} {\enquote {\bibinfo {title} {Bifurcations, and temporal
  and spatial patterns of a modidied {Lotka}-{Volterra} model},}\ }\href
  {\doibase 10.1142/S0218127408021671} {\bibfield  {journal} {\bibinfo
  {journal} {Int. J. Bifurc. Chaos}\ }\textbf {\bibinfo {volume} {18}},\
  \bibinfo {pages} {2223--2248} (\bibinfo {year} {2008})}\BibitemShut {NoStop}%
\bibitem [{\citenamefont {Hu}\ \emph {et~al.}(2022)\citenamefont {Hu},
  \citenamefont {Amor}, \citenamefont {Barbier}, \citenamefont {Bunin},\ and\
  \citenamefont {Gore}}]{Hu22}%
  \BibitemOpen
  \bibfield  {author} {\bibinfo {author} {\bibfnamefont {J.}~\bibnamefont
  {Hu}}, \bibinfo {author} {\bibfnamefont {D.~R.}\ \bibnamefont {Amor}},
  \bibinfo {author} {\bibfnamefont {M.}~\bibnamefont {Barbier}}, \bibinfo
  {author} {\bibfnamefont {G.}~\bibnamefont {Bunin}}, \ and\ \bibinfo {author}
  {\bibfnamefont {J.}~\bibnamefont {Gore}},\ }\bibfield  {title} {\enquote
  {\bibinfo {title} {Emergent phases of ecological diversity and dynamics
  mapped in microcosms},}\ }\href {\doibase 10.1126/science.abm7841} {\bibfield
   {journal} {\bibinfo  {journal} {Science}\ }\textbf {\bibinfo {volume}
  {378}},\ \bibinfo {pages} {85--89} (\bibinfo {year} {2022})}\BibitemShut
  {NoStop}%
\bibitem [{\citenamefont {Roy}\ \emph {et~al.}(2020)\citenamefont {Roy},
  \citenamefont {Barbier}, \citenamefont {Biroli},\ and\ \citenamefont
  {Bunin}}]{Roy20}%
  \BibitemOpen
  \bibfield  {author} {\bibinfo {author} {\bibfnamefont {F.}~\bibnamefont
  {Roy}}, \bibinfo {author} {\bibfnamefont {M.}~\bibnamefont {Barbier}},
  \bibinfo {author} {\bibfnamefont {G.}~\bibnamefont {Biroli}}, \ and\ \bibinfo
  {author} {\bibfnamefont {G.}~\bibnamefont {Bunin}},\ }\bibfield  {title}
  {\enquote {\bibinfo {title} {Complex interactions can create persistent
  fluctuations in high-diversity ecosystems},}\ }\href {\doibase
  10.1371/journal.pcbi.1007827} {\bibfield  {journal} {\bibinfo  {journal}
  {PLOS Computational Biology}\ }\textbf {\bibinfo {volume} {16}},\ \bibinfo
  {pages} {e1007827} (\bibinfo {year} {2020})}\BibitemShut {NoStop}%
\bibitem [{\citenamefont {Marcus}, \citenamefont {Turner},\ and\ \citenamefont
  {Bunin}(2024)}]{Mar24}%
  \BibitemOpen
  \bibfield  {author} {\bibinfo {author} {\bibfnamefont {S.}~\bibnamefont
  {Marcus}}, \bibinfo {author} {\bibfnamefont {A.~M.}\ \bibnamefont {Turner}},
  \ and\ \bibinfo {author} {\bibfnamefont {G.}~\bibnamefont {Bunin}},\
  }\bibfield  {title} {\enquote {\bibinfo {title} {Local and extensive
  fluctuations in sparsely interacting ecological communities},}\ }\href
  {\doibase 10.1103/PhysRevE.109.064410} {\bibfield  {journal} {\bibinfo
  {journal} {Physical Review E}\ }\textbf {\bibinfo {volume} {109}},\ \bibinfo
  {pages} {064410} (\bibinfo {year} {2024})}\BibitemShut {NoStop}%
\bibitem [{\citenamefont {Dalmedigos}\ and\ \citenamefont
  {Bunin}(2020)}]{Dal20}%
  \BibitemOpen
  \bibfield  {author} {\bibinfo {author} {\bibfnamefont {I.}~\bibnamefont
  {Dalmedigos}}\ and\ \bibinfo {author} {\bibfnamefont {G.}~\bibnamefont
  {Bunin}},\ }\bibfield  {title} {\enquote {\bibinfo {title} {Dynamical
  persistence in high-diversity resource-consumer communities},}\ }\href
  {\doibase 10.1371/journal.pcbi.1008189} {\bibfield  {journal} {\bibinfo
  {journal} {PLOS Computational Biology}\ }\textbf {\bibinfo {volume} {16}},\
  \bibinfo {pages} {e1008189} (\bibinfo {year} {2020})}\BibitemShut {NoStop}%
\bibitem [{\citenamefont {Hastings}\ and\ \citenamefont
  {Higgins}(1994)}]{Has94}%
  \BibitemOpen
  \bibfield  {author} {\bibinfo {author} {\bibfnamefont {A.}~\bibnamefont
  {Hastings}}\ and\ \bibinfo {author} {\bibfnamefont {K.}~\bibnamefont
  {Higgins}},\ }\bibfield  {title} {\enquote {\bibinfo {title} {Persistence of
  {{Transients}} in {{Spatially Structured Ecological Models}}},}\ }\href
  {\doibase 10.1126/science.263.5150.1133} {\bibfield  {journal} {\bibinfo
  {journal} {Science}\ }\textbf {\bibinfo {volume} {263}},\ \bibinfo {pages}
  {1133--1136} (\bibinfo {year} {1994})}\BibitemShut {NoStop}%
\bibitem [{\citenamefont {Schreiber}(2003)}]{Schr03}%
  \BibitemOpen
  \bibfield  {author} {\bibinfo {author} {\bibfnamefont {S.~J.}\ \bibnamefont
  {Schreiber}},\ }\bibfield  {title} {\enquote {\bibinfo {title} {Allee
  effects, extinctions, and chaotic transients in simple population models},}\
  }\href {\doibase 10.1016/S0040-5809(03)00072-8} {\bibfield  {journal}
  {\bibinfo  {journal} {Theoretical Population Biology}\ }\textbf {\bibinfo
  {volume} {64}},\ \bibinfo {pages} {201--209} (\bibinfo {year}
  {2003})}\BibitemShut {NoStop}%
\bibitem [{\citenamefont {Morozov}, \citenamefont {Banerjee},\ and\
  \citenamefont {Petrovskii}(2016)}]{Mor16}%
  \BibitemOpen
  \bibfield  {author} {\bibinfo {author} {\bibfnamefont {A.~{\relax Yu}.}\
  \bibnamefont {Morozov}}, \bibinfo {author} {\bibfnamefont {M.}~\bibnamefont
  {Banerjee}}, \ and\ \bibinfo {author} {\bibfnamefont {S.~V.}\ \bibnamefont
  {Petrovskii}},\ }\bibfield  {title} {\enquote {\bibinfo {title} {Long-term
  transients and complex dynamics of a stage-structured population with time
  delay and the {{Allee}} effect},}\ }\href {\doibase
  10.1016/j.jtbi.2016.02.016} {\bibfield  {journal} {\bibinfo  {journal}
  {Journal of Theoretical Biology}\ }\textbf {\bibinfo {volume} {396}},\
  \bibinfo {pages} {116--124} (\bibinfo {year} {2016})}\BibitemShut {NoStop}%
\bibitem [{\citenamefont {Hastings}\ \emph {et~al.}(2018)\citenamefont
  {Hastings}, \citenamefont {Abbott}, \citenamefont {Cuddington}, \citenamefont
  {Francis}, \citenamefont {Gellner}, \citenamefont {Lai}, \citenamefont
  {Morozov}, \citenamefont {Petrovskii}, \citenamefont {Scranton},\ and\
  \citenamefont {Zeeman}}]{Has18}%
  \BibitemOpen
  \bibfield  {author} {\bibinfo {author} {\bibfnamefont {A.}~\bibnamefont
  {Hastings}}, \bibinfo {author} {\bibfnamefont {K.~C.}\ \bibnamefont
  {Abbott}}, \bibinfo {author} {\bibfnamefont {K.}~\bibnamefont {Cuddington}},
  \bibinfo {author} {\bibfnamefont {T.}~\bibnamefont {Francis}}, \bibinfo
  {author} {\bibfnamefont {G.}~\bibnamefont {Gellner}}, \bibinfo {author}
  {\bibfnamefont {Y.-C.}\ \bibnamefont {Lai}}, \bibinfo {author} {\bibfnamefont
  {A.}~\bibnamefont {Morozov}}, \bibinfo {author} {\bibfnamefont
  {S.}~\bibnamefont {Petrovskii}}, \bibinfo {author} {\bibfnamefont
  {K.}~\bibnamefont {Scranton}}, \ and\ \bibinfo {author} {\bibfnamefont
  {M.~L.}\ \bibnamefont {Zeeman}},\ }\bibfield  {title} {\enquote {\bibinfo
  {title} {Transient phenomena in ecology},}\ }\href {\doibase
  10.1126/science.aat6412} {\bibfield  {journal} {\bibinfo  {journal}
  {Science}\ }\textbf {\bibinfo {volume} {361}},\ \bibinfo {pages} {eaat6412}
  (\bibinfo {year} {2018})}\BibitemShut {NoStop}%
\bibitem [{\citenamefont {Sol{\'e}}\ \emph {et~al.}(2002)\citenamefont
  {Sol{\'e}}, \citenamefont {Levin}, \citenamefont {Sol{\'e}}, \citenamefont
  {Alonso},\ and\ \citenamefont {McKane}}]{Sol02}%
  \BibitemOpen
  \bibfield  {author} {\bibinfo {author} {\bibfnamefont {R.~V.}\ \bibnamefont
  {Sol{\'e}}}, \bibinfo {author} {\bibfnamefont {S.~A.}\ \bibnamefont {Levin}},
  \bibinfo {author} {\bibfnamefont {R.~V.}\ \bibnamefont {Sol{\'e}}}, \bibinfo
  {author} {\bibfnamefont {D.}~\bibnamefont {Alonso}}, \ and\ \bibinfo {author}
  {\bibfnamefont {A.}~\bibnamefont {McKane}},\ }\bibfield  {title} {\enquote
  {\bibinfo {title} {Self--organized instability in complex ecosystems},}\
  }\href {\doibase 10.1098/rstb.2001.0992} {\bibfield  {journal} {\bibinfo
  {journal} {Philosophical Transactions of the Royal Society of London. Series
  B: Biological Sciences}\ }\textbf {\bibinfo {volume} {357}},\ \bibinfo
  {pages} {667--681} (\bibinfo {year} {2002})}\BibitemShut {NoStop}%
\bibitem [{\citenamefont {Mora}\ and\ \citenamefont {Bialek}(2011)}]{Mor11}%
  \BibitemOpen
  \bibfield  {author} {\bibinfo {author} {\bibfnamefont {T.}~\bibnamefont
  {Mora}}\ and\ \bibinfo {author} {\bibfnamefont {W.}~\bibnamefont {Bialek}},\
  }\bibfield  {title} {\enquote {\bibinfo {title} {Are {{Biological Systems
  Poised}} at {{Criticality}}?}}\ }\href {\doibase 10.1007/s10955-011-0229-4}
  {\bibfield  {journal} {\bibinfo  {journal} {Journal of Statistical Physics}\
  }\textbf {\bibinfo {volume} {144}},\ \bibinfo {pages} {268--302} (\bibinfo
  {year} {2011})}\BibitemShut {NoStop}%
\bibitem [{\citenamefont {Goel}, \citenamefont {Maitra},\ and\ \citenamefont
  {Montroll}(1971)}]{Goe71}%
  \BibitemOpen
  \bibfield  {author} {\bibinfo {author} {\bibfnamefont {N.~S.}\ \bibnamefont
  {Goel}}, \bibinfo {author} {\bibfnamefont {S.~C.}\ \bibnamefont {Maitra}}, \
  and\ \bibinfo {author} {\bibfnamefont {E.~W.}\ \bibnamefont {Montroll}},\
  }\bibfield  {title} {\enquote {\bibinfo {title} {On the {{Volterra}} and
  {{Other Nonlinear Models}} of {{Interacting Populations}}},}\ }\href
  {\doibase 10.1103/RevModPhys.43.231} {\bibfield  {journal} {\bibinfo
  {journal} {Reviews of Modern Physics}\ }\textbf {\bibinfo {volume} {43}},\
  \bibinfo {pages} {231--276} (\bibinfo {year} {1971})}\BibitemShut {NoStop}%
\bibitem [{\citenamefont {Allesina}\ and\ \citenamefont {Tang}(2015)}]{All15}%
  \BibitemOpen
  \bibfield  {author} {\bibinfo {author} {\bibfnamefont {S.}~\bibnamefont
  {Allesina}}\ and\ \bibinfo {author} {\bibfnamefont {S.}~\bibnamefont
  {Tang}},\ }\bibfield  {title} {\enquote {\bibinfo {title} {The
  stability--complexity relationship at age 40: A random matrix perspective},}\
  }\href {\doibase 10.1007/s10144-014-0471-0} {\bibfield  {journal} {\bibinfo
  {journal} {Population Ecology}\ }\textbf {\bibinfo {volume} {57}},\ \bibinfo
  {pages} {63--75} (\bibinfo {year} {2015})}\BibitemShut {NoStop}%
\bibitem [{\citenamefont {Akjouj}\ \emph {et~al.}(2024)\citenamefont {Akjouj},
  \citenamefont {Barbier}, \citenamefont {Clenet}, \citenamefont {Hachem},
  \citenamefont {Ma{\"i}da}, \citenamefont {Massol}, \citenamefont {Najim},\
  and\ \citenamefont {Tran}}]{Akj24}%
  \BibitemOpen
  \bibfield  {author} {\bibinfo {author} {\bibfnamefont {I.}~\bibnamefont
  {Akjouj}}, \bibinfo {author} {\bibfnamefont {M.}~\bibnamefont {Barbier}},
  \bibinfo {author} {\bibfnamefont {M.}~\bibnamefont {Clenet}}, \bibinfo
  {author} {\bibfnamefont {W.}~\bibnamefont {Hachem}}, \bibinfo {author}
  {\bibfnamefont {M.}~\bibnamefont {Ma{\"i}da}}, \bibinfo {author}
  {\bibfnamefont {F.}~\bibnamefont {Massol}}, \bibinfo {author} {\bibfnamefont
  {J.}~\bibnamefont {Najim}}, \ and\ \bibinfo {author} {\bibfnamefont {V.~C.}\
  \bibnamefont {Tran}},\ }\bibfield  {title} {\enquote {\bibinfo {title}
  {Complex systems in ecology: A guided tour with large {{Lotka}}--{{Volterra}}
  models and random matrices},}\ }\href {\doibase 10.1098/rspa.2023.0284}
  {\bibfield  {journal} {\bibinfo  {journal} {Proceedings of the Royal Society
  A: Mathematical, Physical and Engineering Sciences}\ }\textbf {\bibinfo
  {volume} {480}} (\bibinfo {year} {2024}),\
  10.1098/rspa.2023.0284}\BibitemShut {NoStop}%
\bibitem [{gam()}]{gamma}%
  \BibitemOpen
  \href@noop {} {}\bibinfo {note} {For a simplified model with a random
  distribution of ${\mathbb A}_{ij}=\pm1$, the ratio of predator-prey pairs
  with ${\mathbb A}_{ij}=- {\mathbb A}_{ji}$ is equal to $(1 -
  \gamma)/2$.}\BibitemShut {Stop}%
\bibitem [{\citenamefont {Ginibre}(1965)}]{Gin65}%
  \BibitemOpen
  \bibfield  {author} {\bibinfo {author} {\bibfnamefont {J.}~\bibnamefont
  {Ginibre}},\ }\bibfield  {title} {\enquote {\bibinfo {title} {Statistical
  {{Ensembles}} of {{Complex}}, {{Quaternion}}, and {{Real Matrices}}},}\
  }\href {\doibase 10.1063/1.1704292} {\bibfield  {journal} {\bibinfo
  {journal} {Journal of Mathematical Physics}\ }\textbf {\bibinfo {volume}
  {6}},\ \bibinfo {pages} {440--449} (\bibinfo {year} {1965})}\BibitemShut
  {NoStop}%
\bibitem [{\citenamefont {Mollison}(1991)}]{Mol91}%
  \BibitemOpen
  \bibfield  {author} {\bibinfo {author} {\bibfnamefont {D.}~\bibnamefont
  {Mollison}},\ }\bibfield  {title} {\enquote {\bibinfo {title} {Dependence of
  epidemic and population velocities on basic parameters},}\ }\href {\doibase
  10.1016/0025-5564(91)90009-8} {\bibfield  {journal} {\bibinfo  {journal}
  {Mathematical Biosciences}\ }\textbf {\bibinfo {volume} {107}},\ \bibinfo
  {pages} {255--287} (\bibinfo {year} {1991})}\BibitemShut {NoStop}%
\bibitem [{\citenamefont {Hatton}\ \emph {et~al.}(2024)\citenamefont {Hatton},
  \citenamefont {Mazzarisi}, \citenamefont {Altieri},\ and\ \citenamefont
  {Smerlak}}]{Hat24}%
  \BibitemOpen
  \bibfield  {author} {\bibinfo {author} {\bibfnamefont {I.~A.}\ \bibnamefont
  {Hatton}}, \bibinfo {author} {\bibfnamefont {O.}~\bibnamefont {Mazzarisi}},
  \bibinfo {author} {\bibfnamefont {A.}~\bibnamefont {Altieri}}, \ and\
  \bibinfo {author} {\bibfnamefont {M.}~\bibnamefont {Smerlak}},\ }\bibfield
  {title} {\enquote {\bibinfo {title} {Diversity begets stability:
  {{Sublinear}} growth and competitive coexistence across ecosystems},}\ }\href
  {\doibase 10.1126/science.adg8488} {\bibfield  {journal} {\bibinfo  {journal}
  {Science}\ }\textbf {\bibinfo {volume} {383}},\ \bibinfo {pages} {eadg8488}
  (\bibinfo {year} {2024})}\BibitemShut {NoStop}%
\bibitem [{\citenamefont {{Aguad{\'e}-Gorgori{\'o}}}\ \emph
  {et~al.}(2025)\citenamefont {{Aguad{\'e}-Gorgori{\'o}}}, \citenamefont
  {Lajaaiti}, \citenamefont {Arnoldi},\ and\ \citenamefont {K{\'e}fi}}]{Agu25}%
  \BibitemOpen
  \bibfield  {author} {\bibinfo {author} {\bibfnamefont {G.}~\bibnamefont
  {{Aguad{\'e}-Gorgori{\'o}}}}, \bibinfo {author} {\bibfnamefont
  {I.}~\bibnamefont {Lajaaiti}}, \bibinfo {author} {\bibfnamefont {J.-F.}\
  \bibnamefont {Arnoldi}}, \ and\ \bibinfo {author} {\bibfnamefont
  {S.}~\bibnamefont {K{\'e}fi}},\ }\bibfield  {title} {\enquote {\bibinfo
  {title} {Unpacking sublinear growth: Diversity, stability and coexistence},}\
  }\href {\doibase 10.1111/oik.10980} {\bibfield  {journal} {\bibinfo
  {journal} {Oikos}\ }\textbf {\bibinfo {volume} {2025}},\ \bibinfo {pages}
  {e10980} (\bibinfo {year} {2025})}\BibitemShut {NoStop}%
\bibitem [{\citenamefont {Sommers}\ \emph {et~al.}(1988)\citenamefont
  {Sommers}, \citenamefont {Crisanti}, \citenamefont {Sompolinsky},\ and\
  \citenamefont {Stein}}]{Som88}%
  \BibitemOpen
  \bibfield  {author} {\bibinfo {author} {\bibfnamefont {H.~J.}\ \bibnamefont
  {Sommers}}, \bibinfo {author} {\bibfnamefont {A.}~\bibnamefont {Crisanti}},
  \bibinfo {author} {\bibfnamefont {H.}~\bibnamefont {Sompolinsky}}, \ and\
  \bibinfo {author} {\bibfnamefont {Y.}~\bibnamefont {Stein}},\ }\bibfield
  {title} {\enquote {\bibinfo {title} {Spectrum of {{Large Random Asymmetric
  Matrices}}},}\ }\href {\doibase 10.1103/PhysRevLett.60.1895} {\bibfield
  {journal} {\bibinfo  {journal} {Physical Review Letters}\ }\textbf {\bibinfo
  {volume} {60}},\ \bibinfo {pages} {1895--1898} (\bibinfo {year}
  {1988})}\BibitemShut {NoStop}%
\bibitem [{\citenamefont {Edelman}, \citenamefont {Kostlan},\ and\
  \citenamefont {Shub}(1994)}]{Ede94}%
  \BibitemOpen
  \bibfield  {author} {\bibinfo {author} {\bibfnamefont {A.}~\bibnamefont
  {Edelman}}, \bibinfo {author} {\bibfnamefont {E.}~\bibnamefont {Kostlan}}, \
  and\ \bibinfo {author} {\bibfnamefont {M.}~\bibnamefont {Shub}},\ }\bibfield
  {title} {\enquote {\bibinfo {title} {How many eigenvalues of a random matrix
  are real?}}\ }\href {\doibase 10.1090/S0894-0347-1994-1231689-0} {\bibfield
  {journal} {\bibinfo  {journal} {Journal of the American Mathematical
  Society}\ }\textbf {\bibinfo {volume} {7}},\ \bibinfo {pages} {247--267}
  (\bibinfo {year} {1994})}\BibitemShut {NoStop}%
\bibitem [{\citenamefont {Edelman}(1997)}]{Ede97}%
  \BibitemOpen
  \bibfield  {author} {\bibinfo {author} {\bibfnamefont {A.}~\bibnamefont
  {Edelman}},\ }\bibfield  {title} {\enquote {\bibinfo {title} {The
  {{Probability}} that a {{Random Real Gaussian Matrix}} has{\emph{k}}{{Real
  Eigenvalues}}, {{Related Distributions}}, and the {{Circular Law}}},}\ }\href
  {\doibase 10.1006/jmva.1996.1653} {\bibfield  {journal} {\bibinfo  {journal}
  {Journal of Multivariate Analysis}\ }\textbf {\bibinfo {volume} {60}},\
  \bibinfo {pages} {203--232} (\bibinfo {year} {1997})}\BibitemShut {NoStop}%
\bibitem [{\citenamefont {Kanzieper}\ and\ \citenamefont
  {Akemann}(2005)}]{Kan05}%
  \BibitemOpen
  \bibfield  {author} {\bibinfo {author} {\bibfnamefont {E.}~\bibnamefont
  {Kanzieper}}\ and\ \bibinfo {author} {\bibfnamefont {G.}~\bibnamefont
  {Akemann}},\ }\bibfield  {title} {\enquote {\bibinfo {title} {Statistics of
  {{Real Eigenvalues}} in {{Ginibre}}'s {{Ensemble}} of {{Random Real
  Matrices}}},}\ }\href {\doibase 10.1103/PhysRevLett.95.230201} {\bibfield
  {journal} {\bibinfo  {journal} {Physical Review Letters}\ }\textbf {\bibinfo
  {volume} {95}},\ \bibinfo {pages} {230201} (\bibinfo {year}
  {2005})}\BibitemShut {NoStop}%
\bibitem [{\citenamefont {Benettin}\ \emph
  {et~al.}(1980{\natexlab{a}})\citenamefont {Benettin}, \citenamefont
  {Galgani}, \citenamefont {Giorgilli},\ and\ \citenamefont
  {Strelcyn}}]{Ben80a}%
  \BibitemOpen
  \bibfield  {author} {\bibinfo {author} {\bibfnamefont {G.}~\bibnamefont
  {Benettin}}, \bibinfo {author} {\bibfnamefont {L.}~\bibnamefont {Galgani}},
  \bibinfo {author} {\bibfnamefont {A.}~\bibnamefont {Giorgilli}}, \ and\
  \bibinfo {author} {\bibfnamefont {J.-M.}\ \bibnamefont {Strelcyn}},\
  }\bibfield  {title} {\enquote {\bibinfo {title} {Lyapunov {{Characteristic
  Exponents}} for smooth dynamical systems and for hamiltonian systems; a
  method for computing all of them. {{Part}} 1: {{Theory}}},}\ }\href {\doibase
  10.1007/BF02128236} {\bibfield  {journal} {\bibinfo  {journal} {Meccanica}\
  }\textbf {\bibinfo {volume} {15}},\ \bibinfo {pages} {9--20} (\bibinfo {year}
  {1980}{\natexlab{a}})}\BibitemShut {NoStop}%
\bibitem [{\citenamefont {Benettin}\ \emph
  {et~al.}(1980{\natexlab{b}})\citenamefont {Benettin}, \citenamefont
  {Galgani}, \citenamefont {Giorgilli},\ and\ \citenamefont
  {Strelcyn}}]{Ben80b}%
  \BibitemOpen
  \bibfield  {author} {\bibinfo {author} {\bibfnamefont {G.}~\bibnamefont
  {Benettin}}, \bibinfo {author} {\bibfnamefont {L.}~\bibnamefont {Galgani}},
  \bibinfo {author} {\bibfnamefont {A.}~\bibnamefont {Giorgilli}}, \ and\
  \bibinfo {author} {\bibfnamefont {J.-M.}\ \bibnamefont {Strelcyn}},\
  }\bibfield  {title} {\enquote {\bibinfo {title} {Lyapunov {{Characteristic
  Exponents}} for smooth dynamical systems and for hamiltonian systems; {{A}}
  method for computing all of them. {{Part}} 2: {{Numerical}} application},}\
  }\href {\doibase 10.1007/BF02128237} {\bibfield  {journal} {\bibinfo
  {journal} {Meccanica}\ }\textbf {\bibinfo {volume} {15}},\ \bibinfo {pages}
  {21--30} (\bibinfo {year} {1980}{\natexlab{b}})}\BibitemShut {NoStop}%
\bibitem [{\citenamefont {Smale}(1976)}]{Sma76}%
  \BibitemOpen
  \bibfield  {author} {\bibinfo {author} {\bibfnamefont {S.}~\bibnamefont
  {Smale}},\ }\bibfield  {title} {\enquote {\bibinfo {title} {On the
  differential equations of species in competition},}\ }\href {\doibase
  10.1007/BF00307854} {\bibfield  {journal} {\bibinfo  {journal} {Journal of
  Mathematical Biology}\ }\textbf {\bibinfo {volume} {3}},\ \bibinfo {pages}
  {5--7} (\bibinfo {year} {1976})}\BibitemShut {NoStop}%
\bibitem [{\citenamefont {Mambuca}, \citenamefont {Cammarota},\ and\
  \citenamefont {Neri}(2022)}]{Mam22}%
  \BibitemOpen
  \bibfield  {author} {\bibinfo {author} {\bibfnamefont {A.~M.}\ \bibnamefont
  {Mambuca}}, \bibinfo {author} {\bibfnamefont {C.}~\bibnamefont {Cammarota}},
  \ and\ \bibinfo {author} {\bibfnamefont {I.}~\bibnamefont {Neri}},\
  }\bibfield  {title} {\enquote {\bibinfo {title} {Dynamical systems on large
  networks with predator-prey interactions are stable and exhibit
  oscillations},}\ }\href {\doibase 10.1103/PhysRevE.105.014305} {\bibfield
  {journal} {\bibinfo  {journal} {Physical Review E}\ }\textbf {\bibinfo
  {volume} {105}},\ \bibinfo {pages} {014305} (\bibinfo {year}
  {2022})}\BibitemShut {NoStop}%
\bibitem [{\citenamefont {Fried}, \citenamefont {Shnerb},\ and\ \citenamefont
  {Kessler}(2017)}]{Fri17}%
  \BibitemOpen
  \bibfield  {author} {\bibinfo {author} {\bibfnamefont {Y.}~\bibnamefont
  {Fried}}, \bibinfo {author} {\bibfnamefont {N.~M.}\ \bibnamefont {Shnerb}}, \
  and\ \bibinfo {author} {\bibfnamefont {D.~A.}\ \bibnamefont {Kessler}},\
  }\bibfield  {title} {\enquote {\bibinfo {title} {Alternative steady states in
  ecological networks},}\ }\href {\doibase 10.1103/PhysRevE.96.012412}
  {\bibfield  {journal} {\bibinfo  {journal} {Physical Review E}\ }\textbf
  {\bibinfo {volume} {96}},\ \bibinfo {pages} {012412} (\bibinfo {year}
  {2017})}\BibitemShut {NoStop}%
\bibitem [{\citenamefont {Rall}\ \emph {et~al.}(2010)\citenamefont {Rall},
  \citenamefont {{Vucic-Pestic}}, \citenamefont {Ehnes}, \citenamefont
  {Emmerson},\ and\ \citenamefont {Brose}}]{Ral10}%
  \BibitemOpen
  \bibfield  {author} {\bibinfo {author} {\bibfnamefont {B.~C.}\ \bibnamefont
  {Rall}}, \bibinfo {author} {\bibfnamefont {O.}~\bibnamefont
  {{Vucic-Pestic}}}, \bibinfo {author} {\bibfnamefont {R.~B.}\ \bibnamefont
  {Ehnes}}, \bibinfo {author} {\bibfnamefont {M.}~\bibnamefont {Emmerson}}, \
  and\ \bibinfo {author} {\bibfnamefont {U.}~\bibnamefont {Brose}},\ }\bibfield
   {title} {\enquote {\bibinfo {title} {Temperature, predator--prey interaction
  strength and population stability},}\ }\href {\doibase
  10.1111/j.1365-2486.2009.02124.x} {\bibfield  {journal} {\bibinfo  {journal}
  {Global Change Biology}\ }\textbf {\bibinfo {volume} {16}},\ \bibinfo {pages}
  {2145--2157} (\bibinfo {year} {2010})}\BibitemShut {NoStop}%
\bibitem [{\citenamefont {Englund}\ \emph {et~al.}(2011)\citenamefont
  {Englund}, \citenamefont {{\"O}hlund}, \citenamefont {Hein},\ and\
  \citenamefont {Diehl}}]{Eng11}%
  \BibitemOpen
  \bibfield  {author} {\bibinfo {author} {\bibfnamefont {G.}~\bibnamefont
  {Englund}}, \bibinfo {author} {\bibfnamefont {G.}~\bibnamefont {{\"O}hlund}},
  \bibinfo {author} {\bibfnamefont {C.~L.}\ \bibnamefont {Hein}}, \ and\
  \bibinfo {author} {\bibfnamefont {S.}~\bibnamefont {Diehl}},\ }\bibfield
  {title} {\enquote {\bibinfo {title} {Temperature dependence of the functional
  response},}\ }\href {\doibase 10.1111/j.1461-0248.2011.01661.x} {\bibfield
  {journal} {\bibinfo  {journal} {Ecology Letters}\ }\textbf {\bibinfo {volume}
  {14}},\ \bibinfo {pages} {914--921} (\bibinfo {year} {2011})}\BibitemShut
  {NoStop}%
\bibitem [{\citenamefont {Parain}\ \emph {et~al.}(2019)\citenamefont {Parain},
  \citenamefont {Rohr}, \citenamefont {Gray},\ and\ \citenamefont
  {Bersier}}]{Par19}%
  \BibitemOpen
  \bibfield  {author} {\bibinfo {author} {\bibfnamefont {E.~C.}\ \bibnamefont
  {Parain}}, \bibinfo {author} {\bibfnamefont {R.~P.}\ \bibnamefont {Rohr}},
  \bibinfo {author} {\bibfnamefont {S.~M.}\ \bibnamefont {Gray}}, \ and\
  \bibinfo {author} {\bibfnamefont {L.-F.}\ \bibnamefont {Bersier}},\
  }\bibfield  {title} {\enquote {\bibinfo {title} {Increased {{Temperature
  Disrupts}} the {{Biodiversity}}--{{Ecosystem Functioning Relationship}}},}\
  }\href {\doibase 10.1086/701432} {\bibfield  {journal} {\bibinfo  {journal}
  {The American Naturalist}\ }\textbf {\bibinfo {volume} {193}},\ \bibinfo
  {pages} {227--239} (\bibinfo {year} {2019})}\BibitemShut {NoStop}%
\end{thebibliography}

%

\clearpage

\section*{Supplementary material}
\subsection{Extreme eigenvalue of a real asymmetric random matrix}
\begin{figure}
 \centering
 \hspace{-5mm}
 \includegraphics[width=.9\columnwidth]{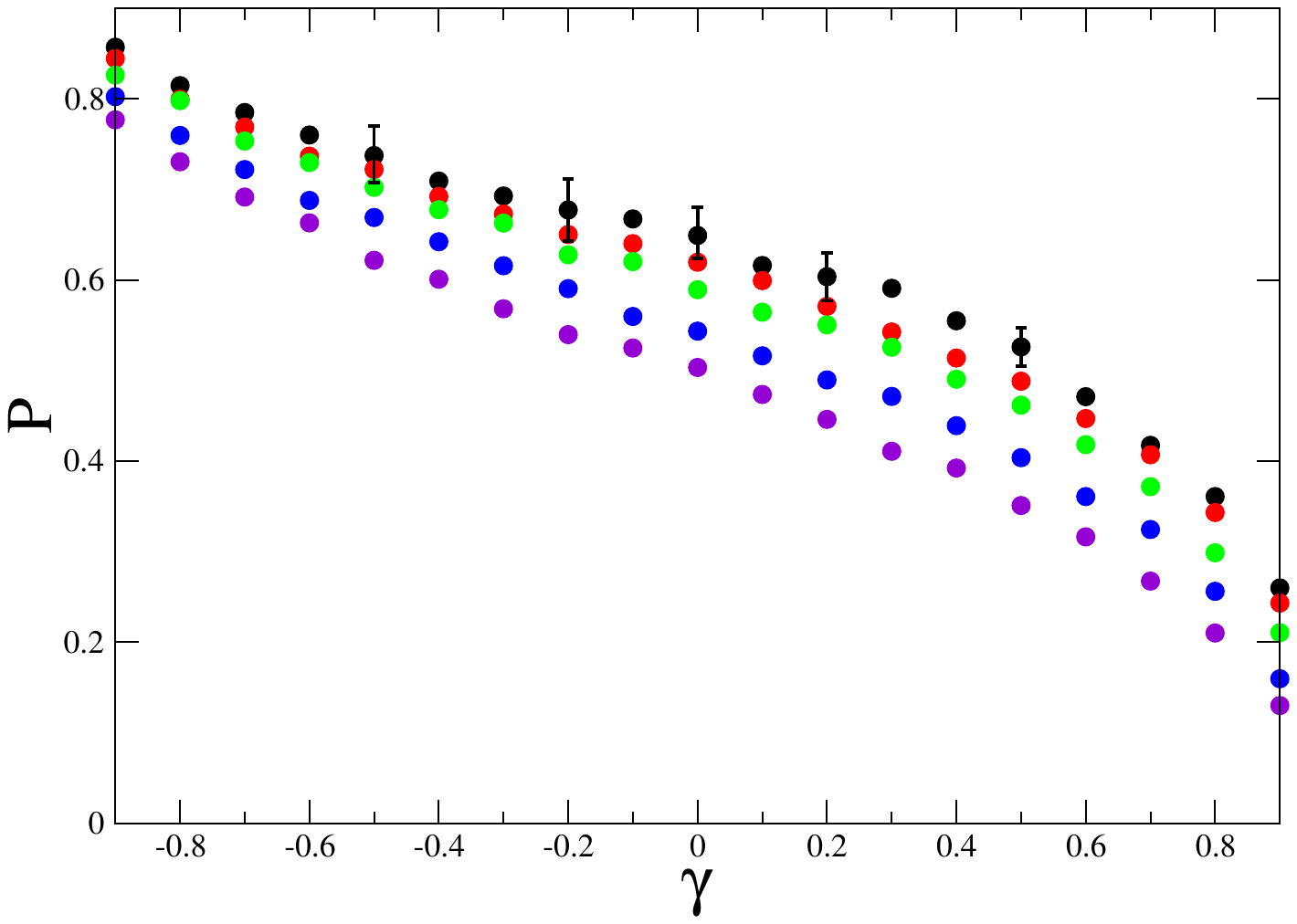}
 \caption{Numerically computed probability that the eigenvalue with largest real part of a random matrix defined by Eq.~(2) is imaginary, as a function of $\gamma$. 
  Data correspond to averages over exact diagonalizations of 10000 matrices of size $S=57$ (violet), 157 (blue) and 557 (green), and of 5000 matrices with $S=1057$ (red) and 2057 (black). 
  Note the limiting cases (not shown) $P(\gamma=-1)=1$ and $P(\gamma=1)=0$.}
 \label{fig:figS1} 
\end{figure}

The occurence of a Hopf bifurcation requires that the first eigenvalue to cross the imaginary axis has a finite imaginary part -- in which case one actually has a pair of different, complex-conjugated eigenvalue. 
Therefore, the probability of a Hopf bifurcation is larger, if the probability that the extreme eigenvalue of the stability matrix has a finite imaginary part is larger. 
Fig.~\ref{fig:figS1} shows this probability for the matrix $\mathbb A$ of Eqs.~(1--2) as a function of $\gamma$ and for different matrix sizes. 
The probability is larger at more negative values of $\gamma$, moreover, it increases with the system size. Once concludes that Hopf bifurcations are the rule rather than the exception for sufficiently large ecosystems with $\gamma <0$.

\subsection{Long-time asymptotic for different initial conditions}
\begin{figure}
 \centering
 \hspace{-5mm}
 \includegraphics[width=\columnwidth]{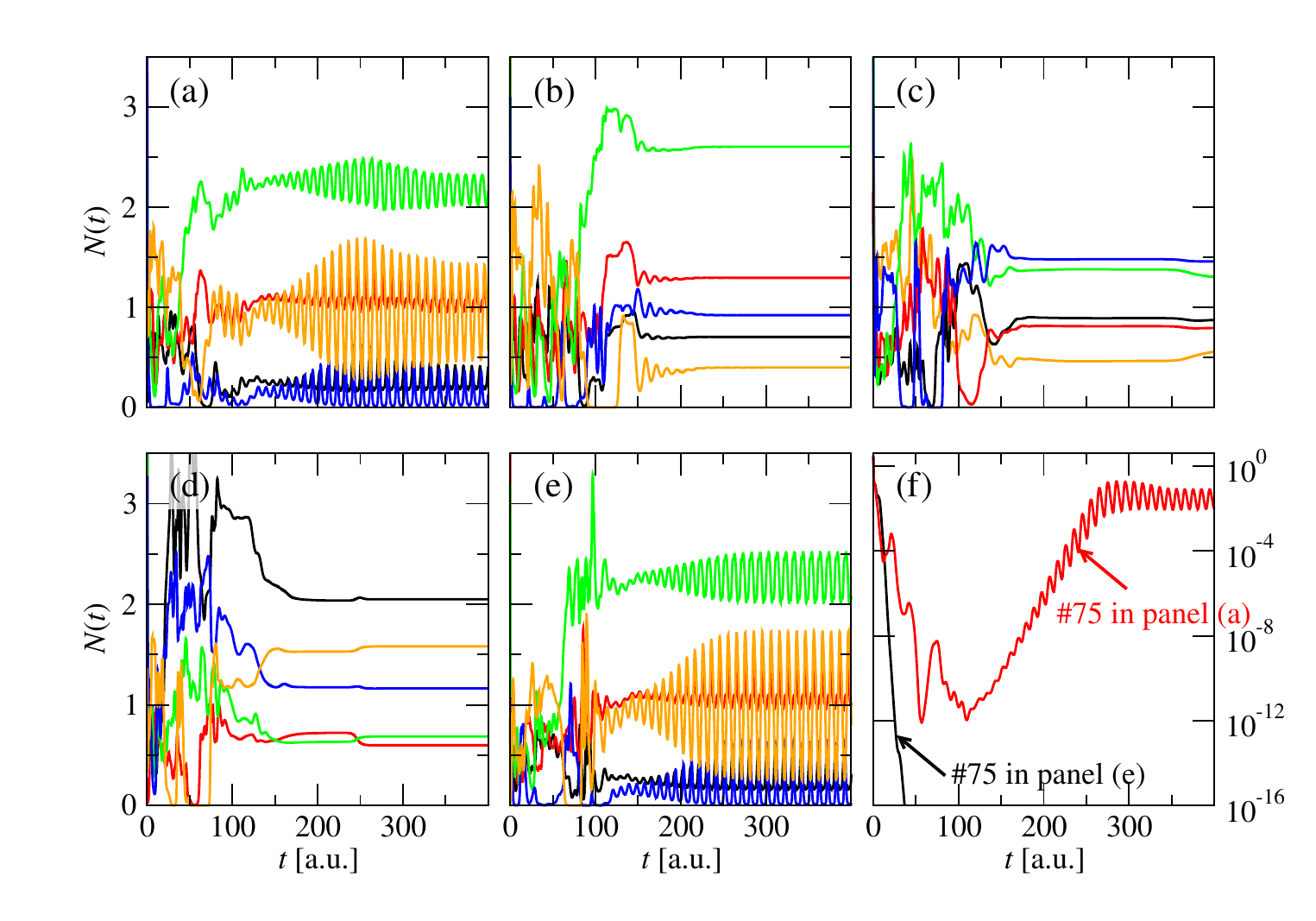}
 \caption{Panels (a-e): Five different dynamics, corresponding to the same realization of the interaction matrix, with five different initial distribution of populations. 
  The limit cycles reached in panels (a) and (e) differ by the presence of one additional species in panel (a). 
  Species oscillate about almost the same average values, with however large amplitude in panel (e). 
  Panel (f): Dynamics of species \#75 which persists in panel (a) but goes extinct in panel (e). 
  Changing the extinction threshold to $N_c=10^{-10}$ leads to two different fixed points for the initial conditions of panels (a) and (e). 
  Five species are shown for each panel.
  The number of surviving species is $N_s=57$ [panel (a)], 51 [panel (b)], 49 [panel (c)], 46 [panel (d)] and 56 [panel (e)].}
 \label{fig:figS2} 
\end{figure}

When $\sigma$ is smaller than the May bound, $\sigma < 1/(1+\gamma)$, the long-time asymptotic is a single, globally attractive fixed point. 
We illustrate that, at larger $\sigma$, and with a finite (though small) extinction threshold, $N_c=10^{-20}$, different asymptotic behaviors can be reached. 
Fig.~\ref{fig:figS2} shows five different dynamics obtained from five different initial distribution of populations subjected to the same interaction matrix ${\mathbb A}_{ij}$. 
For clarity, we show only five species for each case. 
Two initial conditions converge toward  a limit cycle and three toward a fixed point. 
The three fixed points are evidently different. 
The two cycles differ mostly by the survival of species \#75 in panel (a) and its extinction in panel (e). 
As a consequence, the oscillations are similar, but with larger amplitude in panel (e). We stress that the dynamics has been investigated for much longer times than shown, and that in all cases, the dynamics remain the same as for $t \gtrsim 350$, in particular with the same amplitudes of oscillations in panels (a) and (e).

Setting a higher extinction threshold $N_c=10^{-10}$ makes species \#75 disappear also in panel (a), however it further modifies the dynamics so that both the cases of panel (a) and (e) converge toward different fixed points (not shown). 
While the specifically followed dynamics depends on $N_c$, the general conclusion that different dynamics can be followed by different initial conditions is general.

\end{document}